\documentclass[12pt]{article}

\AtBeginDvi{\special{dviout -y=A4P !m dl}}

\usepackage{amsmath}
\usepackage{amssymb}
\usepackage{bm}
\usepackage{mathbbol,bbm}
\usepackage{amsfonts}
\usepackage[latin2]{inputenc}
\usepackage[mathscr]{eucal}
\usepackage{amsthm}
\usepackage{array}
\usepackage{color}
\usepackage[pdftex]{graphicx}
\usepackage{here}

\theoremstyle{definition} 
\newtheorem{lemma}{Lemma}[section]
\newtheorem{definition}[lemma]{Definition}
\newtheorem{theorem}[lemma]{Theorem}
\newtheorem{remark}[lemma]{Remark}
\newtheorem{example}[lemma]{Example}
\newtheorem{corollary}[lemma]{Corollary}
\newtheorem{proposition}[lemma]{Proposition}

\newcommand{\THEN}{\ensuremath{\,\Rightarrow\,}}
 
\newcommand{\A}{\ensuremath{{\mathcal{A}}}}
\newcommand{\B}{\ensuremath{{\mathcal{B}}}}
\newcommand{\C}{\ensuremath{{\mathcal{C}}}}
\newcommand{\X}{\ensuremath{{\mathcal{X}}}}
\renewcommand{\S}{\ensuremath{{\mathcal{S}}}}
 
\newcommand{\Z}{\ensuremath{{\mathbb{Z}}}} 

\newcommand{\re}{\ensuremath{{\mathbb{R}}}} 

\renewcommand{\epsilon}{\varepsilon}
\renewcommand{\phi}{\varphi}

\DeclareMathOperator{\Tr}{Tr}
\DeclareMathOperator*{\argmax}{arg\,max}

\newcommand{\oneover}[1]{\frac{1}{#1}}

 \newcommand{\inner}[2]{%
 \left\langle #1, #2 \right\rangle}
 
\def\s{\mbox{ }}
\def\ds{\s\s}

\def\vfi{\varphi}
\def\N{\mathbb{N}}
\def\R{\mathbb{R}}
\def\iH{\mathcal{H}}
\def\hil{\mathcal{H}}
\def\half{\frac{1}{2}}
\def\bz{\left(}
\def\jz{\right)}
\def\prob{\pi}
\def\state{\vartheta}

\def\egy{\mathbf{\mathrm{1}}}
\def\fact{\eta}
\def\Aset{A\bz \vec{\rho},\vec{\sigma}\jz}
 
\newcommand{\D}[1]{\hat{#1}}
\newcommand{\ki}[1]{{\it #1}}
\newcommand{\kii}[1]{{\it #1}}
\newcommand{\kiii}[1]{#1}

\newcommand{\pmin}[3]{P_{\mathrm{min}}(#1 : #2\,|#3)}
\newcommand{\pn}[3]{P_{T_n}(#1 : #2\,|#3)}

\newcommand{\sr}[2]{S\bz #1\,||\,#2\jz}
\newcommand{\srm}[2]{S_{\mathrm{M}}\bz #1\,||\,#2\jz}
\newcommand{\cb}[2]{C\bz #1,\,#2\jz}
\newcommand{\cbm}[2]{C_{\mathrm{M}}\bz #1,\,#2\jz}
\newcommand{\hli}[1]{\underline B\bz #1| \vec\rho\, ||\, \vec\sigma\jz}
\newcommand{\hls}[1]{\overline B\bz #1| \vec\rho \,||\, \vec\sigma\jz}
\newcommand{\hlim}[1]{B\bz #1| \vec\rho \,||\, \vec\sigma\jz}
\newcommand{\derleft}[1]{\partial^{-} #1}
\newcommand{\derright}[1]{\partial^{+} #1}
\newcommand{\I}[1]{I_{#1}}

\newcommand{\unit}{\Eins}
\newcommand{\proj}[1]{|#1\rangle\langle #1|}
\newcommand{\vect}[1]{\underline{#1}}

\newcommand{\factstates}[1]{\S_{\mathrm{fact}}(#1)}

\DeclareMathOperator{\supp}{supp}

\renewcommand\theenumi{(\arabic{enumi})}

\begin{document}

\centerline{\huge Error exponents in hypothesis testing}
\medskip

 \centerline{\huge for correlated states
on a spin chain}
\bigskip
\s

\bigskip

 \centerline{\large Fumio Hiai\footnote{Electronic mail: hiai@math.is.tohoku.ac.jp}, 
Mil\'an Mosonyi\footnote{Electronic mail: milan.mosonyi@gmail.com}}
  \medskip

\centerline{\textit{Graduate School of Information Sciences, Tohoku University}}

\centerline{\textit{Aoba-ku, Sendai 980-8579, Japan}}
\bigskip

\centerline{\large Tomohiro Ogawa\footnote{Electronic mail: ogawa@quantum.jst.go.jp}}
\medskip

\centerline{\textit{PRESTO, Japan Science and Technology Agency}}

\centerline{\textit{4-1-8 Honcho Kawaguchi, Saitama, 332-0012, Japan}}
\bigskip
\s

\medskip

\begin{abstract}
 We study various error exponents in a binary hypothesis testing problem
 and extend recent results on the quantum Chernoff and Hoeffding bounds for product states to a
 setting when both the null-hypothesis and the counter-hypothesis can be correlated
 states on a spin chain. Our results apply to states satisfying a certain factorization property; typical examples are the global Gibbs states of
 translation-invariant finite-range interactions as well as certain finitely correlated states.
\end{abstract}
\medskip


\section{Introduction}
  
We study the asymptotics of the error probabilities in a binary
hypothesis testing problem for quantum systems. In a rather general
setting (used generally in the information-spectrum approach
\cite{Han, NH}),
one can consider a sequence of finite-level quantum systems with
(finite-dimensional) Hilbert spaces
$\vec\hil=\{\hil_n\}_{n=1}^{\infty}$. Assume that we know a priori that the
$n$th system is in state $\rho_n$ (null-hypothesis $H_0$)
or in state $\sigma_n$ (counter-hypothesis $H_1$).
The hypothesis testing problem for the $n$th system is to decide between the above two
options, based on the outcome of a binary measurement on the system. 

A measurement in our setting means a binary positive operator valued measure
$\{T_n,I_{n}-T_n\}$ where $0\le T_n\le I_{n}$ corresponds to outcome $0$ and
$I_{n}-T_n$ to outcome $1$. If the outcome of the measurement is $0$ (resp.~$1$) then
hypothesis $H_0$ (resp.~$H_1$) is accepted. Obviously we can identify the measurement
with the single operator $T_n$. An erroneous decision is made if $H_1$ ($H_0$) is
accepted when the true state of the system is $\rho_n$ ($\sigma_n$); the probabilities
of these events are the error probabilities of the first (second) kinds, given by
\begin{equation*}
\alpha_n (T_n) := \rho_n(I_n-T_n)=\Tr\D{\rho}_n(I_{n} -T_n)\ds\ds\text{ and }\ds\ds
\beta_n (T_n) :=  \sigma_n(T_n)=\Tr\D{\sigma}_nT_n\,,
\end{equation*}
respectively. (Here $\D{\omega}$ denotes the density operator of a state $\omega$,
given by $\omega(A)=\Tr \D{\omega}A,\s A\in\B(\hil_n)$.)

Apart from the trivial case when $\supp\D{\rho}_n\perp\supp\D{\sigma}_n$, one cannot
find a measurement making both error probabilities to vanish; in general, there is a
tradeoff between the two. 
In the general cases of interest the error probabilities are expected to tend to zero
asymptotically (typically with an exponential speed) when the measurements $T_n$ are
chosen in an optimal way.
In the asymmetric setting of \ki{Stein's lemma}
\cite{DZ,HP-1}
the exponential decay of the $\beta_n$'s is studied either under the constraint that
the $\alpha_n$'s tend to $0$, or that the $\alpha_n$'s stay under a constant bound.
As it was shown in \cite{HP-1} and \cite{ON}, in the i.i.d.~case (i.e.\ when
$\iH_n=\iH_1^{\otimes n}$, $\rho_n=\rho_1^{\otimes n}$ and
$\sigma_n=\sigma_1^{\otimes n}$)
the optimal exponential decay rate is given by $-\sr{\rho_1}{\sigma_1}$, the negative
relative entropy of $\rho_1$ and $\sigma_1$, thus giving an operational interpretation
to relative entropy. This result was later extended to cases when the sequences
$\vec\rho:=\{\rho_n\}_{n=1}^{\infty}$ and $\vec\sigma:=\{\sigma_n\}_{n=1}^{\infty}$
consist of restrictions of an ergodic state $\rho$ and a shift-invariant product state
$\sigma$ on a spin chain \cite{HP,BS}.
In the symmetric setting of the \ki{Chernoff bound} \cite{Aud,ANSzV,NSz} the exponential
decay of the average of the two error probabilities is of interest. As it was shown in
\cite{Aud} and \cite{NSz}, the best exponential decay rate in the i.i.d.~case is given
by $-\cb{\rho_1}{\sigma_1}$, with
\begin{equation}\label{psi}
\cb{\rho_1}{\sigma_1}:=-\min_{0\le s\le 1}\psi(s)\,,\ds\ds\ds
\psi(s):=\log\Tr \D{\rho}_1^s\D{\sigma}_1^{1-s}\,.
\end{equation}
The above result shows that the quantity $C$ plays a similar role in symmetric
hypothesis testing as the relative entropy does in the asymmetric case. 


When an exponential bound is given on the decay of the $\alpha_n$'s,
our interest lies in the following quantities \cite{OH,NH}:
\begin{eqnarray}
\hli{r}&:=&\inf_{\{T_n\}}\biggl\{
\liminf_{n\rightarrow\infty} \oneover{n}\log\beta_n(T_n)
\biggm|
\limsup_{n\rightarrow\infty} \oneover{n}\log\alpha_n(T_n) < -r
\biggr\}\,,\label{Hb1}\\
\hls{r}&:=&\inf_{\{T_n\}}\biggl\{
\limsup_{n\rightarrow\infty} \oneover{n}\log\beta_n(T_n)
\biggm|
\limsup_{n\rightarrow\infty} \oneover{n}\log\alpha_n(T_n) < -r
\biggr\}\,,\label{Hb2}\\
\hlim{r}&:=&\inf_{\{T_n\}}\biggl\{
\lim_{n\rightarrow\infty} \oneover{n}\log\beta_n(T_n)
\biggm|
\limsup_{n\rightarrow\infty} \oneover{n}\log\alpha_n(T_n) < -r
\biggr\}\,.\label{Hb3}
\end{eqnarray}
Based on the techniques developed in \cite{Aud} and \cite{NSz}, the identity
\begin{equation}\label{Hoeffding}
\hls{r}=\hlim{r}=-b(r)\,,\ds\ds\ds b(r):=\max_{0\le s<1}\frac{-sr-\psi(s)}{1-s}
\end{equation}
was proven for $0<r\le \sr{\rho_1}{\sigma_1}$ in the i.i.d.~setting in \cite{Hayashi}
(where $\hls{r}\le -b(r)$ was shown) and \cite{Nagaoka} (where the inequality
$\hlim{r}\ge -b(r)$ was provided), thus establishing the theorem for the quantum
\ki{Hoeffding bound}. 
 
In this paper we will mainly consider the situation when $\rho_n$ and $\sigma_n$ are the
$n$-site restrictions of states $\rho$ and $\sigma$ on an infinite spin chain,
satisfying a certain factorization property. Typical examples of such states are the
global Gibbs states of translation-invariant finite-range interactions \cite{HMO} and
certain finitely correlated states \cite{FNW,HMO}.
Our main result is that (\ref{Hoeffding}) holds for such states 
when $\psi$ in \eqref{psi} is replaced with 
\begin{equation*}
\psi(s):=\lim_n\frac{1}{n}\log\Tr\D{\rho}_n^s\D{\sigma}_n^{1-s}\,,\ds\ds\ds s\in [0,1]\,.
\end{equation*}
As a side-result, we recover the quantum Chernoff bound (already proven in \cite{HMO})
and a Stein-type upper bound for states of the above type.

\section{Preliminaries and upper bounds}\label{Preliminaries section}
\subsection{Error exponents: upper bounds}\label{error exponents}
Let $A$ and $B$ be nonnegative operators on a finite-dimensional Hilbert space $\hil$.
It is easy to see that 
\begin{equation} 
\min_{0\le T\le I}\{\Tr A(I-T) +\Tr BT \}
=\oneover{2}\Tr(A+B)-\oneover{2}\Tr|A-B|\,,
\label{tracenorm}
\end{equation}
and the minimum is attained at $\{A-B>0\}$, the spectral projection of $A-B$
corresponding to the positive part of the spectrum.
The following was shown in \cite{Aud}:
\begin{lemma}\label{Audenaert}
Let $A$ and $B$ be nonnegative operators on a finite-dimensional Hilbert space $\hil$.
Then
\begin{equation}\label{trace inequality}
 \oneover{2}\Tr(A+B)-\oneover{2}\Tr|A-B| \le \Tr A^{s}B^{1-s}\,,
\qquad 0\le s\le 1.
\end{equation}
\end{lemma}
All along the paper we use the convention $0^s:=0,\s s\in\R$; in particular, $A^0$ and
$B^0$ are defined to be the support projections of $A$ and $B$, respectively. With this
convention $s\mapsto\Tr A^{s}B^{1-s}$ is a continuous function on $\R$. In \ref{min} we 
mention another representation of the quantity \eqref{tracenorm} given in \cite{Yuen}.
\smallskip

Consider now the hypothesis testing problem described in the Introduction, and assume
that we know a priori that the $n$th system is in the state $\rho_n$ with probability
$\prob_n\in (0,1)$ or in the state $\sigma_n$ with probability $1-\prob_n$. Then the
Bayesian probability of an erroneous decision based on the test $T_n$ is
\begin{equation*} 
\pn{\rho_n}{\sigma_n}{\prob_n}
:=\prob_n\,\alpha_n(T_n)+(1-\prob_n)\,\beta_n(T_n)\,,
\end{equation*}
and by applying Lemma \ref{Audenaert} to $A:=\prob_n\D{\rho}_n$ and
$B:=(1-\prob_n)\D{\sigma}_n$ we get
\begin{equation}\label{minimum error}
\pmin{\rho_n}{\sigma_n}{\prob_n}
:=\min_{0\le T_n\le I_n}\{\pn{\rho_n}{\sigma_n}{\pi_n}\}
\le \prob_n^{s}(1-\prob_n)^{1-s}\,\Tr \D{\rho}_n^{s}\D{\sigma}_n^{1-s}\,.
\end{equation}
Note that the optimal test is of the form
$\{\prob_n\D{\rho}_n-(1-\prob_n)\D{\sigma}_n>0\}$.
Let
\begin{equation}\label{def:psin}
\psi_n(s):=\frac{1}{n}\log\Tr \D{\rho}_n^{s}\D{\sigma}_n^{1-s}\,,\ds\ds\ds s\in\R\,.
\end{equation}
It is easy to see that $\psi_n$ is a convex function on $\R$ for all $n\in\N$. Next, let
\begin{equation*}
\psi(s):=\limsup_{n\to\infty}\psi_n(s)\,,\ds\ds\ds s\in [0,1]\,,
\end{equation*}
and
\begin{equation}\label{def:vfi}
\vfi(a):=\sup_{0\le s\le 1}\{as-\psi(s)\}\,,\ds\ds\ds a\in\R\,,
\end{equation}
be its \ki{conjugate} (or \ki{polar}) function.
If $\prob_n=\prob$ is independent of $n$ then (\ref{minimum error}) implies 
\begin{equation}\label{Chernoff1}
\limsup_{n\to\infty}\frac{1}{n}\log \pmin{\rho_n}{\sigma_n}{\prob}
\le \inf_{0\le s\le 1}\psi(s)=-\vfi(0)\,,
\end{equation}
as it was pointed out in the i.i.d.~case in \cite{Aud}.
\smallskip

After this preparation, we prove the following:
\begin{lemma}
\label{direct-part}
Let $\D{\rho}_n$ and $\D{\sigma}_n$
be density operators on a Hilbert space $\hil_n$ for each $n\in\N$.
Then for any $a\in\R$
\begin{equation}\label{Chernoff upper}
\limsup_{n\to\infty}\frac{1}{n}\log\min_{0\le T_n\le I_n}
\{e^{-na}\alpha_n(T_n)+\beta_n(T_n)\}\le 
-\vfi(a)\,.
\end{equation}
Moreover, for $S_{n,a}:=\{e^{-na}\D{\rho}_n-\D{\sigma}_n>0\}$ we have
\begin{align*}
\limsup_{n\rightarrow\infty}\oneover{n}\log\alpha_n(S_{n,a}) &\le -\{\phi(a)-a\} \,,
\\
\limsup_{n\rightarrow\infty}\oneover{n}\log\beta_n(S_{n,a}) &\le -\phi(a)\,.
\end{align*}
\end{lemma}
\begin{proof}
Consider formula (\ref{minimum error}) with $\prob_n:=\frac{e^{-na}}{1+e^{-na}}$ for
a fixed $a\in\R$. The optimal test is then
$\{\prob_n\D{\rho}_n-(1-\prob_n)\D{\sigma}_n>0\}=S_{n,a}$, and
by multiplying (\ref{minimum error}) by $1+e^{-na}$ we get
\begin{equation*}
\min_{0\le T_n\le I_n}\{e^{-na}\alpha_n(T_n)+ \beta_n(T_n)\}
=e^{-na}\alpha_n(S_{n,a})+ \beta_n(S_{n,a})\le 
e^{-nas}\Tr \D{\rho}_n^{s}\D{\sigma}_n^{1-s}\,,
\end{equation*}
which implies
\begin{equation*}
\alpha_n(S_{n,a}) \le e^{na} e^{-nas}\Tr \D{\rho}_n^{s}\D{\sigma}_n^{1-s},\ds\ds\ds
\beta_n(S_{n,a}) \le e^{-nas}\Tr\D{\rho}_n^{s}\D{\sigma}_n^{1-s}
\end{equation*}
for any $s\in[0,1]$.
Thus the statement follows.
\end{proof}

Consider now an asymmetric hypothesis testing problem with an exponential bound on the
decay of the $\alpha_n$'s. The relevant error exponents in this case are given in
\eqref{Hb1}, \eqref{Hb2} and \eqref{Hb3}.
Obviously for any fixed $r\in\R$
\begin{equation}\label{inequality:exponents}
\hli{r}\le\hls{r}\le\hlim{r}\,,
\end{equation}
and all the above quantities are monotonically increasing functions of $r$. Note that
for $r<0$ the choice $T_n:=I_n$ yields $\hli{r}=\hls{r}=\hlim{r}=-\infty$, hence the
above quantities are only interesting for $r\ge 0$.

Lemma \ref{direct-part} yields the following corollary, that can be considered as the
direct part of the theorem for the quantum Hoeffding bound:
\begin{corollary}\label{cor direct-part}
In the above setting
\begin{equation*}
\hls{r}
\le -\sup_{a:\,\phi(a)-a> r}\phi(a)\,.
\end{equation*}
\end{corollary}

\medskip

The converse part of \eqref{Chernoff1}, inequality
\begin{equation}\label{Chernoff2}
\liminf_{n\to\infty}\frac{1}{n}\log \pmin{\rho_n}{\sigma_n}{\prob}
\ge \inf_{0\le s\le 1}\psi(s)=-\vfi(0)
\end{equation}
was shown in the i.i.d.~setting in \cite{NSz}. Inequalities \eqref{Chernoff1} and
\eqref{Chernoff2} together give the theorem for the quantum \ki{Chernoff bound} in the
i.i.d.\ case.
The main idea in proving the lower bound is to 
relate the problem to the classical hypothesis testing problem of a certain pair of
classical probability measures associated to the original pair of quantum states.
The same method was used to prove the lower bound in the theorem for the quantum
Hoeffding bound in \cite{Nagaoka}. 
In Section \ref{converse-part 1} we follow
(a slight modification of) this method to show that 
the converse part of inequality \eqref{Chernoff upper} in Lemma \ref{direct-part} holds (in the above general setting) if the functions $\psi_n$ converge to a differentiable function on $\R$. Apart from yielding the lower bound in the i.i.d.~setting as a special case, there are examples for correlated states on a spin chain for which this criterion can be verified (see Example \ref{example 1}). In general, however, differentiabilty seems to be rather difficult to prove, therefore we follow a different approach in Section \ref{converse-part 2} to obtain the
converse part for a certain class of states on a spin chain, which we introduce in Section \ref{spin chains}.
 
\subsection{Spin chains and factorization property}\label{spin chains}

Let $\iH$ be a finite-dimensional Hilbert space and $\A\subset\B(\iH)$ be a
$C^*$-subalgebra. 
For all $k,l\in\Z,\s k\le l$, the finite-size algebra
$\C_{[k,l]}:=\otimes_{k\le i\le l}\A$ is naturally embedded into all $\C_{[m,n]}$ with
$m\le k, n\ge l$, hence one can define
$\C_{\mathrm{loc}}:=\bigcup_{k,l\in\N}\C_{[k,l]}$, which is a pre-$C^*$-algebra with
unit $\unit$. The \ki{spin chain} $\C$ with one-site algebra $\A$ is the uniform closure
of $\C_{\mathrm{loc}}$. It is natural to consider $\C$ as the infinite tensor power of
$\A$, hence the notation $\C=\otimes_{k\in\Z}\A$ is also used. The right shift
automorphism $\gamma$ is the unique extension of the maps
$\gamma_{kl}:\,\C_{[k,l]}\to\C_{[k,l+1]},\ds X\mapsto\unit_{\A}\otimes X$.

States on the spin chain are positive linear functionals on $\C$ that take the value
$1$ on $\unit$. A state $\omega$ is \ki{translation-invariant} if
$\omega\circ\gamma=\omega$ holds. A translation-invariant state $\omega$ is uniquely
determined by $\vec\omega:=\{\omega_n\}_{n=1}^{\infty}$, where $\omega_n$ is its
restriction onto $\C_{[1,n]}$.
\begin{definition}
\label{def:factorization}
A translation-invariant state $\omega$ on the spin chain
satisfies upper/lower factorization properties
if there exists a positive constant $\fact\in\re$ such that
\begin{align}
\omega &\le \fact\,\omega_{(-\infty,0]}\otimes\omega_{[1,+\infty)}
\quad \text{(upper factorization)},
\label{upper}
\\
\omega &\ge \fact^{-1}\,\omega_{(-\infty,0]}\otimes\omega_{[1,+\infty)}
\quad \text{(lower factorization)}.
\label{lower}
\end{align}
\end{definition}
\smallskip

For a fixed $m\in\N$ any number $n\in\N$ can be written in the form $n=km+r$ with
$k,r\in\N,\s 1\le r\le m$, and consecutive applications of the above inequalities
give
\begin{align}
\omega_{[1,n]} &\le \fact^{k}\,\omega_{[1,m]}^{\otimes k}\otimes\omega_{[1,r]}
\quad \text{(upper factorization)},\label{upper2}
\\
\omega_{[1,n]} &\ge \fact^{-k}\,\omega_{[1,m]}^{\otimes k}\otimes\omega_{[1,r]}
\quad \text{(lower factorization)}.\label{lower2}
\end{align}
On the other hand, it is easily seen by taking $n=2m$ for an arbitrarily large $m$
that inequalities \eqref{upper2} and \eqref{lower2} imply \eqref{upper} and
\eqref{lower}, respectively. We will
use the notation $\vec{\rho}\in\factstates{\C}$ 
if $\rho_n,\,n\in\N$, are the $n$-site restrictions of a shift-invariant state
$\rho$ on $\C$ that satisfies the factorization properties above.

Obviously, a product state $\omega:=\omega_1^{\otimes\infty}$ satisfies both upper and
lower factorization properties. As it was shown in \cite{HMO}, finitely correlated
states \cite{FNW} satisfy upper factorization property, and in some special cases
(e.g.\ locally faithful Markov states) also lower factorization property
\cite{HMO,HMOP}. 
Another important class of states that satisfy both upper and lower factorization
properties is that of the global Gibbs states of translation-invariant finite-range
interactions. This result was also shown in \cite{HMO}, based on the perturbation bounds
developed in \cite{LRB}. 

Now let $\vec{\rho},\vec{\sigma}\in\factstates{\C}$.
Without loss of generality we can assume that they have the same factorization constant
$\fact$. 
If $\supp\D{\rho}_m\perp\supp\D{\sigma}_m$ for some $m$ then the upper factorization
property \eqref{upper} yields that $\supp\D{\rho}_n\perp\supp\D{\sigma}_n$ for all
$n\in\N$. Since in this case the  hypothesis testing problem is trivial, we will
always assume that $\supp\D{\rho}_n$ and $\supp\D{\sigma}_n$ are not orthogonal to
each other for any $n$, as far as the case
$\vec\rho,\vec\sigma\in\factstates{\C}$ is concerned.
  
Let $\psi_n\,,n\in\N$, be as given in \eqref{def:psin}.
The following lemma was shown in \cite{HMO}; for readers' convenience and because we
need an intermediate formula for later purposes, we give a detailed proof here.
\begin{lemma}\label{factorization limit}
The limit $\psi(s):=\lim_{n\rightarrow\infty}\psi_n(s)$ exists for all $s\in [0,1]$,
and $\psi$ is a convex and continuous function on $[0,1]$.
\end{lemma}
\begin{proof}
Let $m\in\N$ be fixed and $n=km+r, \s 1\le r\le m$.
Upper factorization property together with
the operator monotonicity of the function $x\mapsto x^{s},\s 0\le s\le 1$,
implies
\begin{align*}
\Tr\D{\rho}_n^{s}\D{\sigma}_n^{1-s}
\le M \fact^k (\Tr\D{\rho}_{m}^{s}\D{\sigma}_{m}^{1-s})^k,
\end{align*}
where $M=\max\{\Tr\D{\rho}_r^{s}\D{\sigma}_r^{1-s} : 1\le r\le m,\, 0\le s\le 1\}$,
and hence
\begin{align*}
\psi_n(s) \le \oneover{n}\log M + \frac{k}{n}\log\fact + \frac{km}{n} \psi_m(s).
\end{align*}
Taking the $\limsup$ in $n$, we obtain
\begin{align*}
\limsup_{n\rightarrow\infty}\psi_n(s) \le \psi_m(s) + \oneover{m}\log\fact.
\end{align*}
Taking the $\liminf$ in $m$ then gives the existence of the limit.
Being the pointwise limit of convex functions, $\psi$ is convex (and hence continuous
in $(0,1)$).
 
In the same way as above, lower factorization property implies
\begin{align*}
\psi_m(s) - \oneover{m}\log\fact
\le \liminf_{n\rightarrow\infty}\psi(s)
\end{align*}
and we obtain the bound
\begin{align}\label{approx-psi2}
\psi_m(s) - \oneover{m}\log\fact
\le \psi(s)
\le \psi_m(s) + \oneover{m}\log\fact
\end{align}
for every $m\in\N$.
This shows that $\psi$ is the uniform limit of the $\psi_n$'s, and hence the continuity
of $\psi$ follows.
\end{proof}

\section{A Chernoff-type theorem}\label{lower bounds}

In this section we complement inequality \eqref{Chernoff upper} of Lemma \ref{direct-part}.
Our main interest is in the situation when $\vec{\rho},\vec{\sigma}\in\factstates{\C}$; we treat this case in Section \ref{converse-part 2}. The main idea in this case is to use the lower factorization property to reduce the problem to the i.i.d.~setting. {\color{black}} In Section \ref{converse-part 1} we prove the converse inequality of \eqref{Chernoff upper} under the assumption that the functions $\psi_n$ defined in \eqref{def:psin} converge to a differentiable function $\psi$ on $\R$. 
Even though this condition may seem to be rather abstract and difficult to verify in general, this approach has at least two merits. First, the converse inequality for the i.i.d.~situation (needed in Section \ref{converse-part 2}) follows as a special case. Second, it provides an extension from the i.i.d.~situation that can be different from requiring the lower factorization property to hold, as we point out in Remark \ref{separation}.

\subsection{Lower bound under differentiability}\label{converse-part 1}

Let $\psi_n,\,n\in\N$, be as given in \eqref{def:psin} and assume that 

\renewcommand\theenumi{(a\arabic{enumi})}
\begin{enumerate}
\item\label{existence}
 the limit $\psi(s):=\lim_n\,\psi_n(s)$ exists as a real number
for all $s\in\R$;
\item\label{diff2}
 $\psi$ is a differentiable function on $\R$.
\end{enumerate}

Note that assumption (a1) implies that $\supp\D{\rho}_n$ cannot be orthogonal
to $\supp\D{\sigma}_n$, except for finitely many $n$'s.
Since all the $\psi_n$'s are convex on $\R$, $\psi$ is a convex
function on $\R$ as well. 
Let $\tilde\psi(s):=\psi(1-s),\,s\in\R$; then
 \begin{equation*}
\psi^*(x):=\sup_{s\in\R}\{sx-\psi(s)\}\ds\ds\text{and}\ds\ds
(\tilde\psi)^*(x):=\sup_{s\in\R}\{sx-\tilde\psi(s)\}=x+\psi^*(-x)
\end{equation*}
are convex functions on $\R$ with values in $(-\infty,+\infty]$ (usually referred to as the \ki{Legendre-Fenchel transforms} of $\psi$ and $\tilde\psi$). 
Let $\vfi(a):=\sup_{0\le s\le 1}\{as-\psi(s)\},\,a\in\R$, as given in \eqref{def:vfi}; then
$\vfi(a)\le \psi^*(a)$ and
$\vfi(a)= \psi^*(a)$ if and only if $\psi'(0)\le a\le\psi'(1)$.

We will use (a slight modification of) the method of \cite{NSz} and \cite{Nagaoka} together with the G\"artner-Ellis
theorem (see e.g.\ \cite[Section 2.3]{DZ}) to show the following:
\begin{theorem}\label{lower bound app}
Under the above assumptions
\begin{equation}\label{statement}
\lim_{n\to\infty}\frac{1}{n}\log\min_{0\le T_n\le I_n}
\{e^{-na}\alpha_n(T_n)+\beta_n(T_n)\}
= -\vfi(a)
\end{equation}
for any $a\in\R$ with $a\ne \psi'(0),\,\psi'(1)$. Moreover, if $\psi^*$ is continuous
at $\psi'(0)$ and at $\psi'(1)$
then \eqref{statement} holds for all $a\in\R$.
\end{theorem}
\begin{proof}
Thanks to Lemma \ref{direct-part} it suffices to prove that 
for any sequence of
tests $\{T_n\}$ we have
\begin{equation}\label{liminf-ineq}
\liminf_{n\to\infty}\frac{1}{n}\log\left\{e^{-na}\alpha_n(T_n)+\beta_n(T_n)\right\}
\ge -\vfi(a).
\end{equation}
 Let 
\begin{equation}\label{decomposition}
\D{\rho}_n=\sum_{i\in I_n}\lambda_i P_i\,,\ds\ds\ds
\D{\sigma}_n=\sum_{j\in J_n}\eta_j Q_j
\end{equation}
be a decomposition of the densities $\D{\rho}_n$ and $\D{\sigma}_n$, where $P_i,Q_j$
are projections and $\lambda_i,\eta_j>0$ for all $i\in I_n,\,j\in J_n$. Define the
classical discrete positive measures on $I_n\times J_n$ by
\begin{equation*}
p_n(i,j):=\lambda_i\Tr P_iQ_j\,,\ds\ds\ds
q_n(i,j):=\eta_j\Tr P_iQ_j\,.
\end{equation*}
Note that $p_n(I_n\times J_n) \le 1$ and $q_n(I_n\times J_n) \le 1$, and
$\supp p_n=\supp q_n$ holds for all $n$. Moreover, it is easy to see that 
$\Tr\D{\rho_n}^s\D{\sigma}_n^{1-s}=\sum_{(i,j)\in I_n\times J_n}p_n(i,j)^s q_n(i,j)^{1-s}$ and 
assumption \ref{existence} implies 
\begin{equation}\label{limit probabilities}
\lim_{n\to\infty}\frac{1}{n}\log p_n\bz I_n\times J_n\jz=\psi(0)\ds\ds\ds\text{and}\ds\ds\ds
\lim_{n\to\infty}\frac{1}{n}\log q_n\bz I_n\times J_n\jz=\psi(1)\,.
\end{equation}

Let $S_{n,a}:=\{e^{-na}\D{\rho}_n-\D{\sigma}_n>0\}$. Then 
\begin{eqnarray*}
e_n(a)&:=&\min_{0\le T_n\le I_n}\{e^{-na}\alpha_n(T_n)+\beta_n(T_n)\}
= e^{-na}\Tr \D{\rho}_n (I-S_{n,a})+ \Tr\D{\sigma}_n S_{n,a}\\
&=&  e^{-na}\sum_{i\in I_n} \lambda_i \Tr (I-S_{n,a}) P_i(I-S_{n,a})
+\sum_{j\in J_n} \eta_j\Tr S_{n,a} Q_j S_{n,a}\\
&\ge& e^{-na}\sum_{(i,j)\in I_n\times J_n} \lambda_i \Tr (I-S_{n,a}) P_i(I-S_{n,a}) Q_j +\sum_{(i,j)\in I_n\times J_n} \eta_j \Tr S_{n,a} Q_j S_{n,a} P_i\\
&\ge& \sum_{(i,j)\in I_n\times J_n} \min\{e^{-na}\lambda_i,\eta_j\}  \Tr \left[(I-S_{n,a}) P_i (I-S_{n,a})+S_{n,a} P_i S_{n,a}\right] Q_j\,.
\end{eqnarray*}
Now by \cite[Lemma 9]{Hayashi2} we have 
\begin{equation*}
\half (I-S_{n,a}) P_i (I-S_{n,a})+\half S_{n,a} P_i S_{n,a}
\ge \frac{1}{4} P_i\,,\ds\ds\ds i\in I_n\,.
\end{equation*}
(This can also be seen from the operator convexity of the function
$f_A:\,X\mapsto X^*AX$ for a positive semidefinite $A$, which fact can be verified by
a straightforward computation \cite[Lemma 5]{OH}.)
As a consequence,
\begin{eqnarray*}
2e_n(a)&\ge&  \sum_{(i,j)\in I_n\times J_n} \min\{e^{-na}\lambda_i,\eta_j\}  \Tr  P_i Q_j=
\sum_{(i,j)\in I_n\times J_n} \min\{e^{-na} p_n(i,j),q_n(i,j)\} \\
&=& e^{-na} p_n\bz \{e^{-na}p_n(i,j)\le q_n(i,j)\} \jz
+ q_n\bz \{e^{-na}p_n(i,j)> q_n(i,j)\} \jz\\
&=&
e^{-na} p_n\bz \{X_n\ge -a\} \jz
 +q_n\bz \{Y_n>a\}\jz\,,
\end{eqnarray*}
where
\begin{equation*}
X_n(i,j):=\frac{1}{n}\log\frac{q_n(i,j)}{p_n(i,j)}\,,\ds\ds\ds
Y_n(i,j):=\frac{1}{n}\log\frac{p_n(i,j)}{q_n(i,j)}
\end{equation*}
are random variables with corresponding distribution measures
$\mu_n^{(1)}:=p_n\circ X_n^{-1}$ and $\mu_n^{(2)}:=q_n\circ Y_n^{-1}$.
A straightforward computation shows that
\begin{equation*}
\log\int e^{nsx}\,d\mu_n^{(1)}=n\psi_n(1-s)\,,\ds\ds\ds
\log\int e^{nsx}\,d\mu_n^{(2)}=n\psi_n(s)\,.
\end{equation*}
Under our assumptions (a1) and (a2), the G\"artner-Ellis theorem yields that
\begin{eqnarray*}
\liminf_n\frac{1}{n}\log p_n\bz \{X_n\ge -a\}\jz&\ge&-\inf_{x> -a} (\tilde\psi)^*(x)\,,\\
\liminf_n\frac{1}{n}\log q_n\bz \{Y_n>a\}\jz&\ge&-\inf_{x> a} \psi^*(x)\,,
\end{eqnarray*}
and therefore
\begin{equation*}
\liminf_n\frac{1}{n}\log e_n(a)\ge -m(a)
:=-\min\left\{\inf_{x> a} \psi^*(x),\,a+\inf_{x> -a} (\tilde\psi)^*(x)\right\}\,.
\end{equation*}
We remark that the G\"artner-Ellis theorem is usually stated for probability measures while our measures $p_n$ and $q_n$ are in general subnormalized. However, this case follows immediately from the standard version due to the existence of the limits in \eqref{limit probabilities}.

Since $\inf_{x\in\R}\psi^*(x)=\psi^*(\psi'(0))=-\psi(0)$, we get
$\inf_{x> a} \psi^*(x)=-\psi(0)$ for 
$a<\psi'(0)$, and in this case also $\vfi(a)=-\psi(0)$, hence $m(a)\le \vfi(a)$.
Moreover, the same holds for $a=\psi'(0)$ if it is a continuity point of $\psi^*$.
Similarly,
$\inf_{x\in\R}(\tilde\psi)^*(x)=(\tilde\psi)^*(\tilde\psi'(0))=-\tilde\psi(0)=-\psi(1)$
implies $\inf_{x>-a} (\tilde\psi)^*(x)=-\psi(1)$ for 
$-a<\tilde\psi'(0)=-\psi'(1)$, hence $m(a)\le a-\psi(1)=\vfi(a)$ when $a>\psi'(1)$.
Again, continuity of $(\tilde\psi)^*$ at $\tilde\psi'(0)$ (i.e.\ continuity of $\psi^*$
at $\psi'(1)$) yields $m(a)\le \vfi(a)$ for  $a=\psi'(1)$. The proof is finished by
noting that for $\psi'(0)<a<\psi'(1)$ we have $\inf_{x> a} \psi^*(x)=\psi^*(a)=\vfi(a)$.
\end{proof}

In the proof of Theorem \ref{Nagaoka-bound} we will need that 
\eqref{statement} holds for every $a\in\R$ in the special case when both $\rho$ and $\sigma$
are shift-invariant product states on a spin chain. To prove this, we first
give the following lemma, which may be interesting by itself:

\begin{lemma}\label{affine structure}
Let $a$ and $b$ positive elements in a $C^*$-algebra $\A\subset\B(\iH)$, where $\iH$ is a finite-dimensional Hilbert space. Let 
$a=\sum_{i\in I} \lambda_i P_i$ and 
$b=\sum_{j\in J} \eta_j Q_j$\s
be their spectral decompositions (with all $\lambda_i,\eta_j>0$), and let $\psi(s):=\log\Tr a^s b^{1-s},\,s\in\R$. Then the following are equivalent: 
\renewcommand\theenumi{(\roman{enumi})}
\begin{enumerate}
\item \label {affine} $\psi$ is an affine function on $\R$;
\item \label{derivative} there exists an $s\in\R$ such that $\psi''(s)=0$;
\item \label{structure} there exist $i_1,\ldots,i_m\in I,\s j_1,\ldots, j_m\in J$ and a $\delta>0$ 
such that
\begin{equation*}
P_{i_k}\vee Q_{j_k}\perp P_{i_l}\vee Q_{j_l}\,,\ds\ds k\neq l\,,
\end{equation*}
\begin{equation*}
P_{i_k}\wedge Q_{j_k}\ne 0\ds\text{ and }\ds
\lambda_{i_k}=\delta\,\eta_{j_k}\,,\ds\ds k=1,\ldots,m\,,
\end{equation*}
 and that $\sum_{i\in I\setminus\{i_i,\ldots,i_m\}} P_i$,\s $\sum_{j\in J\setminus\{j_i,\ldots,j_m\}} Q_j$ and 
$\sum_{k=1}^m\,P_{i_k}\vee Q_{j_k}$ are mutually orthogonal projections.
\end{enumerate}
Moreover, if $\A$ is isomorphic to the function algebra on a finite set $\X$ then the above are also equivalent to
\begin{enumerate}
\addtocounter{enumi}{3}
\item\label{classical}
There exists a $\delta>0$ such that $a(x)=\delta\,b(x)$ for every $x\in\X$ with $a(x)b(x)\ne 0$.
\end{enumerate}
\end{lemma}
\begin{proof}
\ref{affine} $\Rightarrow$ \ref{derivative} is obvious, and \ref{structure} $\Rightarrow$ \ref{affine} is easy to check. 

To see \ref{derivative} $\Rightarrow$ \ref{structure},
define a function $f$ and probability distributions $p_s,\,s\in\R$, on $I\times J$ by
\begin{equation*}
f(i,j):=\log\lambda_i-\log\eta_j\,,\ds\ds\ds\ds\ds
p_s(i,j):=\frac{\lambda_i^s\eta_j^{1-s}\,\Tr P_iQ_j}{\Tr A^s B^{1-s}}\,.
\end{equation*}
Note that the support of $p_s$ is the same for all $s\in\R$.
A straightforward computation yields that  
\begin{equation*}
\psi''(s)=\sum_{ij}f(i,j)^2 p_s(i,j)-\big(\sum_{ij}f(i,j) p_s(i,j)\big)^2\,,
\end{equation*}
which is $0$ if and only if $f$ is constant on the support of $p_s$, i.e.~there exists a constant $c\in\R$ such that 
$\log\lambda_i-\log\eta_j=c$ or equivalently $\lambda_i=\delta\,\eta_j$  with $\delta:=e^{c}$ for all $i,j$ such that $\Tr P_i Q_j\ne 0$. Now if both  $\Tr P_i Q_j\ne 0$ and $\Tr P_{i'} Q_j\ne 0$ then $\lambda_i=\delta\,\eta_j=\lambda_{i'}$, hence $i=i'$. The same way $\Tr P_i Q_j\ne 0$ and $\Tr P_{i} Q_{j'}\ne 0$
imply $j=j'$, and the rest of the statement follows.

Finally, the equivalence of \ref{classical} and \ref{structure} in the commutative case is easy to see.
\end{proof}

\begin{corollary}\label{iid lower bound}
If $\rho$ and $\sigma$ are shift-invariant product states then \eqref{statement} holds
for all $a\in\R$.
\end{corollary}
\begin{proof}
First note that assumptions (a1) and (a2) are satisfied in this case. The
spectral decompositions $\D{\rho}_1=\sum_{i\in I_1}\lambda_i P_i$ and
$\D{\sigma}_1=\sum_{j\in J_1}\eta_j Q_j$ induce decompositions
$\D{\rho}_n=\sum_{\vect{i}\in I_1^n}\lambda_{\vect{i}} P_{\vect{i}}$ and
$\D{\sigma}_n=\sum_{\vect{j}\in J_1^n}\eta_{\vect{j}} Q_{\vect{j}}$
as in \eqref{decomposition} for all $n\in\N$, 
where $\lambda_{\vect{i}}:=\lambda_{i_1}\cdot\ldots\cdot \lambda_{i_n}$,\s 
$P_{\vect{i}}:=P_{i_1}\otimes\ldots\otimes P_{i_n}$ and similarly for 
$\eta_{\vect{j}}$ and $Q_{\vect{j}}$.
%
%
As a consequence, for the associated classical probabilities we have
$p_n=p_1^{\otimes n},\,q_n=q_1^{\otimes n}$. Then
$\psi(s)=\log\sum_{(i,j)\in I_1\times J_1}\, p_1(i,j)^s q_1(i,j)^{1-s}$ and 
\begin{equation}\label{eq1}
\liminf_n\frac{1}{n}\log e_n(a)\ge
\liminf_n\frac{1}{n}\log\sum_{(\vect{i},\vect{j})\in I_1^n\times J_1^n}
\min\{e^{-na}p_1^{\otimes n}(\vect{i},\vect{j}),\,q_1^{\otimes n}(\vect{i},\vect{j})\}\,.
\end{equation}
Now we distinguish two cases. If $\psi$ is not affine then by (ii) of Lemma
\ref{affine structure} we have $\psi''(s)>0$ for all $s\in\R$, and this implies that
$\psi'(s)$ is in the interior of $\{x\in\R:\psi^*(x)<+\infty\}$ for all $s\in\R$.
As a consequence, $\psi^*$ is continuous at $\psi'(0)$ and at $\psi'(1)$,
and this case is covered by Theorem \ref{lower bound app}.

Assume now that $\psi$ is affine; then by (iv) of Lemma \ref{affine structure} we
have $p_1(i,j)=\delta\,q_1(i,j)$ for all $(i,j)\in S:=\supp p_1\cap\supp q_1$, hence
\begin{equation*}
\psi(s)=\log\sum_{(i,j)\in S} \bz\delta\,q_1(i,j)\jz^s q_1(i,j)^{1-s}=s\log\delta+\log q_1(S)\,,
\end{equation*}
and therefore
\begin{equation}\label{eq2}
\vfi(a)=\begin{cases}
-\log q_1(S)\,,& a\le \log \delta\,,\\
-\log q_1(S)+a-\log \delta\,, & a>\log\delta\,.
\end{cases}
\end{equation}
On the other hand,
\begin{equation*}
\min\{e^{-na}p_1^{\otimes n}(\vect{i},\vect{j}),\,q_1^{\otimes n}(\vect{i},\vect{j})\}
= \begin{cases}
0\,,& (\vect{i},\vect{j})\notin S^n\,,\\
q_1^{\otimes n}(\vect{i},\vect{j})\,, & (\vect{i},\vect{j})\in S^n,\s a\le \log\delta\,,\\
e^{-na}\delta^n\,q_1^{\otimes n}(\vect{i},\vect{j})\,,& (\vect{i},\vect{j})\in S^n,\s a> \log\delta\,,
\end{cases}
\end{equation*}
and thus
\begin{equation}\label{eq3}
\frac{1}{n}\log\sum_{(\vect{i},\vect{j})\in I_1^n\times J_1^n}\min\{e^{-na}p_1^{\otimes n}(\vect{i},\vect{j}),\,q_1^{\otimes n}(\vect{i},\vect{j})\}
=\begin{cases}
\log q_1(S)\,,& a\le \log\delta\,,\\
\log q_1(S)-a+\log\delta\,,& a> \log\delta\,.
\end{cases}
\end{equation}
Formulas \eqref{eq1}, \eqref{eq2} and \eqref{eq3} together give the desired statement.
\end{proof}
 
\subsection{Lower bound under factorization}
\label{converse-part 2}

Assume now that $\vec{\rho},\vec{\sigma}\in\factstates{\C}$. Then the limit
$\psi(s):=\lim_n\psi_n(s)$ exists for all $s\in[0,1]$ and $\psi$ is continuous on $[0,1]$, as was shown
in Lemma \ref{factorization limit}. 
For each $m\in\N$ let
\begin{equation*}
\vfi_m(a):=\max_{0\le s\le 1}\{as-\psi(s)\}\,,\ds\ds\ds a\in\R
\end{equation*}
be the polar function of $\psi_m$. The bound
(\ref{approx-psi2}) implies that 
\begin{align}
\phi_m(a) -\oneover{m}\log\fact
\le \phi(a)
\le \phi_m(a) +\oneover{m}\log\fact
\label{approx-phi2}
\end{align}
for every $a\in\R$ and $m\in\N$, hence $\vfi$ is the uniform limit of the sequence
$\{\vfi_m\}$.
\begin{theorem}
\label{Nagaoka-bound}
Let $\vec{\rho},\vec{\sigma}\in\factstates{\C}$. Then for any $a\in\R$
\begin{equation}\label{statement2}
\lim_{n\rightarrow\infty}\frac{1}{n}\log
\min_{0\le T_n\le I_n}\{e^{-na}\alpha_n(T_n)+\beta_n(T_n)\} = -\phi(a).
\end{equation}
\end{theorem}
\begin{proof}
Due to Lemma \ref{direct-part} it suffices to prove that \eqref{liminf-ineq} holds
for any sequence of
tests $\{T_n\}$.
We can assume that $\rho$ and $\sigma$ have the same factorization constant $\eta$.
Let $m\in\N$ be fixed and write $n>m$ in the form $n=(k-1)m+r$ with $1\le r\le m$. 
With $\gamma:=\min\{1,\,e^{(m-r)a}\,;\,0\le r<m\}$ we have
\begin{align}\label{lowerfact bound}
e^{-na}\alpha_n(T_n)+\beta_n(T_n)
&\ge
\gamma\bz e^{-kma}\alpha_n(T_n)+\beta_n(T_n)\jz
\nonumber \\
&=
\gamma\bz e^{-kma}\alpha_{km}(T_n\otimes I_{[n+1,km]})
+\beta_{km}(T_n\otimes I_{[n+1,km]})\jz
\nonumber \\
&\ge
\gamma\,\fact^{-(k-1)} \min_{0\le T\le I}\left\{e^{-kma}
\Tr[\D{\rho}_m^{\otimes k}(I-T)]+\Tr[\D{\sigma}_m^{\otimes k}T ]
\right\},
\end{align}
where we used lower factorization property in the last step.
Let
\begin{align*}
\Psi_m(s):=\log\Tr\D{\rho}_m^{s}\D{\sigma}_m^{1-s}=m\psi_m(s)\ds\text{ and }\ds \Phi_m(a):=\max_{0\le s\le 1}\left\{as-\Psi_m(s)\right\}=m\vfi_m(a/m)\,.
\end{align*}
By Corollary \ref{iid lower bound} we have
\begin{align*}
\liminf_{k\rightarrow\infty}\oneover{k}\log\min_{0\le T\le I}\left\{
 e^{-kma}\Tr[\D{\rho}_m^{\otimes k}(I-T)]+\Tr[\D{\sigma}_m^{\otimes k}T]
\right\}
\ge -\Phi_m(ma).
\end{align*}
Combining it with \eqref{lowerfact bound} we obtain
\begin{eqnarray*}
\liminf_{n\rightarrow\infty}
\oneover{n}\log\left\{e^{-na}\alpha_n(T_n)+ \beta_n(T_n)\right\}
\nonumber 
&\ge&
-\oneover{m}\log\fact-\oneover{m}\Phi_m(ma)=-\oneover{m}\log\fact-\phi_m(a)\\
&\ge&
-\oneover{m}\log\fact-\phi(a) -\oneover{m}\log\fact,
\end{eqnarray*}
where we used \eqref{approx-phi2} in the last inequality.
Taking the limit in $m$ gives the assertion.
\end{proof}

\subsection{Some remarks}\label{some remarks}

Note that equations \eqref{statement} and \eqref{statement2} can be reformulated as 
\begin{equation}\label{Chernoff type}
\lim_{n\to\infty}\frac{1}{n}\log\pmin{\rho_n}{\sigma_n}{\prob_n}=
-\max\{\vfi(a),\vfi(a)-a\}=
\begin{cases}
-\vfi(a)& \text{ if }  a\ge 0\,,\\
-\vfi(a)+a & \text{ if }  a<0\,, 
\end{cases}
\end{equation}
with $\pi_n:=\frac{e^{-na}}{1+e^{-na}}$, therefore giving an extension of the theorem for the Chernoff bound to a setting when the prior probabilities are not constant, but depend on $n$ in the given way. 
From Lemma \ref{direct-part} we get that \eqref{Chernoff type} holds whenever $a\in\Aset$, where $\Aset$ denotes the set of all $a\in\R$ for which inequality \eqref{liminf-ineq} is satisfied. 
In particular, if $0\in\Aset$ then we recover the theorem for the Chernoff bound
 \cite{ANSzV,HMO}. This is the case e.g.~when $\vec{\rho},\vec{\sigma}\in\factstates{\C}$ (since $\Aset=\R$ by Theorem \ref{Nagaoka-bound}) and when assumptions \ref{existence} and \ref{diff2} are satisfied and $0\ne \derright{\psi}(0)$ and $0\ne \derleft{\psi}(1)$ (since $\Aset\supset \R\setminus\{\derright{\psi}(0),\derleft{\psi}(1)\}$ by Theorem \ref{lower bound app}).

Alternatively, one can interpret Theorems \ref{lower bound app} and \ref{Nagaoka-bound} as the theorem for the Chernoff bound in the setting when hypothesis testing is performed between the states $\sigma_n$ and the unnormalized states $e^{-na}\rho_n$. Indeed, in the setting of Section \ref{converse-part 1} or Section \ref{converse-part 2} we have 
\begin{equation*}
\psi_a(s):=\lim_n\frac{1}{n}\log\Tr \bz e^{-na}\D{\rho}_n\jz^s \D{\sigma}_n^{1-s}=-\{as-\psi(s)\}\,,\ds\ds s\in [0,1]\,,
\end{equation*}
and
\begin{equation*}
\lim_n\frac{1}{n}\log\pmin{e^{-na}\rho_n}{\sigma_n}{\pi}=\min_{0\le s\le 1}\psi_a(s)
\end{equation*}
for any $\pi\in (0,1)$ and $a\in\Aset$.

\section{The Hoeffding bound and related exponents}\label{main results}

Our main goal in this section is to derive the theorem for the Hoeffding bound in the settings of Sections \ref{converse-part 1} and \ref{converse-part 2}. To treat the two settings in a unified way, we derive all our results under the following assumptions:
 
\bigskip\noindent
\renewcommand\theenumi{(A\arabic{enumi})}
\begin{enumerate}
\item \label{existence2}
The limit $\psi(s):=\lim_{n\to\infty}\psi_n(s)$ exists
as a real number for all $s\in[0,1]$ and $\psi$ is continuous on $[0,1]$.
\item \label{lower bound2}
The inequality
\begin{equation}\label{assump-ineq}
\lim_{n\to\infty}\frac{1}{n}\log\min_{0\le T_n\le I_n}
\{e^{-na}\alpha_n(T_n)+\beta_n(T_n)\} \ge -\vfi(a)
\end{equation}
holds for all $a<\derleft{\psi}(1)$ (the left derivative of $\psi$ at $1$), except possibly for finitely many values of $a$,
where $\vfi$ is given in \eqref{def:vfi}.
\end{enumerate}

Though assumptions \ref{existence2} and \ref{lower bound2} are admittedly rather artificial, they have the merits that they are satisfied in the cases of our interest on the one hand (see Section \ref{lower bounds}), and on the other hand they give the minimal requirements under which the results of this section are valid, thus providing a better view on the logical relations among our results.
%

We begin by introducing $\tilde\psi(s):=\psi(1-s),\, s\in [0,1]$ and its polar function
\begin{equation*}
\tilde\vfi(a):=\max_{0\le s\le 1}\{as-\tilde\psi(s)\}=a+\vfi(-a), \ds\ds\ds a\in\R.
\end{equation*}
We also define
\begin{equation}
\hat\vfi(a):=\tilde\vfi(-a)=\vfi(a)-a\,,\ds\ds\ds a\in\R\,.
\end{equation}
In \ref{graphs} we give an illustration of the above definitions and the properties listed in the following:
\renewcommand\theenumi{(\roman{enumi})}
\begin{lemma} \label{lem:phi}
The functions $\vfi$ and $\hat\vfi$ have the following properties:
\begin{enumerate}
\item $\phi$ is convex, continuous, and increasing on $\re$.
Moreover, it is constant $-\psi(0)$ on the interval $\left(-\infty,\derright{\psi}(0)\right]$ and
strictly increasing on $\left(\derright{\psi}(0),+\infty\right)$.
\item $\hat\vfi$ is convex, continuous, and decreasing on $\re$.
Moreover, it is strictly decreasing on the interval $\left( -\infty,\derleft{\psi}(1)\right)$ and is constant $-\psi(1)$ on the interval
$\left[\derleft{\psi}(1),+\infty\right)$.
\end{enumerate}
\end{lemma}
\begin{proof}
All properties follow immediately from the very definitions of $\vfi$ and $\hat\vfi$, except for strict monotonicity. We only prove it for $\hat\vfi$, as the proof for $\vfi$ is completely similar.
Note that
\begin{align*}
\hat\phi(a)=\phi(a)-a=\max_{0\le s\le 1}\{a(s-1)-\psi(s)\},
\end{align*}
and let $s_a:=\argmax_{0\le s\le 1}\{a(s-1)-\psi(s)\}$.
It follows from $a(1-1)-\psi(1)=-\psi(1)$ that if $a$ is such that $\phi(a)-a>-\psi(1)$ (i.e.~$a<\derleft{\psi}(1)$) then $s_a<1$.
Hence for $b<a$
\begin{align*}
\phi(a)-a=a(s_a-1)-\psi(s_a)<b(s_a-1)-\psi(s_a)\le\phi(b)-b.
\end{align*}
\end{proof}
\begin{remark}\label{infinity}
Due to the above listed properties, one can extend $\vfi$ and $\hat\vfi$ to continuous and monotonic functions on $\R\cup\{+\infty\}$ by defining $\vfi(+\infty):=+\infty$ and $\hat\vfi(+\infty):=-\psi(1)$.
\end{remark}
\begin{remark}\label{rem:properties}
Note that $\psi_n(s)\le 0$ for all $s\in [0,1]$, and $n\in\N$, hence the same holds for $\psi$.
If $\supp\D{\rho}_n\le\supp\D{\sigma}_n$ then $\psi_n(1)=\frac{1}{n}\log\Tr\left[\D{\rho}_n\supp\D{\sigma}_n\right]=0$, and a straightforward computation shows that
\begin{equation*}
\derleft{\psi}_n(1)=\frac{1}{n}\sr{\rho_n}{\sigma_n}\,,
\end{equation*}
where 
$\sr{\rho_n}{\sigma_n}:=\Tr \D{\rho}_n(\log\D{\rho}_n-\log\D{\sigma}_n)$ is the \ki{relative entropy} of the states $\rho_n$ and $\sigma_n$. 
Convexity of $\psi_n$ implies
\begin{equation}\label{convexity}
\psi_n(s)\ge \psi_n(1)+(s-1)\,\derleft{\psi}_n(1)=(s-1)\,\frac{1}{n}\sr{\rho_n}{\sigma_n}\,,\ds\ds\ds s\in [0,1].
\end{equation}
Assume now that $\supp\D{\rho}_n\le\supp\D{\sigma}_n$ holds for all large enough $n$, and that 
the mean relative entropy
\begin{equation*}
\srm{\vec{\rho}}{\vec{\sigma}}:=\lim_{n\rightarrow\infty}\oneover{n}\sr{\rho_n}{\sigma_n}
\end{equation*}
exists. (Note that if $\vec{\rho},\vec{\sigma}\in\factstates{\C}$ then 
if $\supp\D{\rho}_m\subset\supp\D{\sigma}_m$ for some $m\in\N$ then it also holds for all $n\in\N$, and 
the mean relative entropy exists even if one requires only the upper factorization property to hold \cite[Theorem 2.1]{HP}.)
Then $\psi(1)=0$, and taking the limit in \eqref{convexity}
yields
\begin{equation*}
\psi(s)\ge (s-1)\,\srm{\rho}{\sigma}\,,\ds\ds\ds s\in [0,1]\,,
\end{equation*}
therefore
\begin{equation}\label{derleft-relentr}
\derleft{\psi}(1)\le \srm{\rho}{\sigma}\,.
\end{equation}
Similarly, if 
we replace the condition $\supp\D{\rho}_n\le\supp\D{\sigma}_n$ with $\supp\D{\rho}_n\ge\supp\D{\sigma}_n$ in the above argument then we get $\psi(0)=0$ and $\derright{\psi}(0)\ge -\srm{\sigma}{\rho}$.

Note that if $\rho_n=\rho_1^{\otimes n}$ and $\sigma_n=\sigma_1^{\otimes n}$, $n\in\N$, with $\supp\D{\rho}_1\wedge\supp\D{\sigma}_1\ne 0$ and $\supp\D{\rho}_1\nleq\supp\D{\sigma}_1$,
then $\derleft{\psi}(1)$ is finite while $\srm{\rho}{\sigma}=+\infty$, hence \eqref{derleft-relentr} cannot be expected to hold as an equality in general.
In \ref{differentiability} we show examples for correlated states $\rho,\sigma$ on a spin chain for which  $\derleft{\psi}(1)= \srm{\vec{\rho}}{\vec{\sigma}}$ can be shown by an explicit computation.
\end{remark}

\begin{lemma}
\label{converse-part}
Let $a<\derleft{\psi}(1)$. Then for any sequence of tests $\{T_n\}$ satisfying
\begin{align*}
\limsup_{n\rightarrow\infty}\oneover{n}\log\alpha_n(T_n)
\le -\{\phi(a)-a\},
\end{align*}
we have
\begin{align*}
\liminf_{n\rightarrow\infty}\oneover{n}\log\beta_n(T_n)
\ge -\phi(a).
\end{align*}
\end{lemma}
\begin{proof}
We follow the same argument as in \cite{Nagaoka}.
For any $b$ satisfying \eqref{assump-ineq} we have
\begin{align*}
-\phi(b)
&\le
\liminf_{n\rightarrow\infty}
\oneover{n}\log\left\{e^{-nb}\alpha_n(T_n)+\beta_n(T_n)\right\}
\nonumber\\
&\le
\max\left\{
\liminf_{n\rightarrow\infty}\oneover{n}\log\beta_n(T_n),
-b+\limsup_{n\rightarrow\infty}\oneover{n}\log\alpha_n(T_n)
\right\}
\nonumber\\
&\le
\max\left\{
\liminf_{n\rightarrow\infty}\oneover{n}\log\beta_n(T_n),
-b-\phi(a)+a
\right\}
\end{align*}
If $a<b<\derleft{\psi}(1)$, then $-\phi(a)+a<-\phi(b)+b$ by Lemma \ref{lem:phi}\,(ii),
hence
\begin{align*}
\liminf_{n\rightarrow\infty}\oneover{n}\log\beta_n(T_n)
\ge -\phi(b)\,.
\end{align*}
Continuity of $\phi$ then yields
\begin{equation*}
\liminf_{n\rightarrow\infty}\oneover{n}\log\beta_n(T_n)
\ge -\phi(a).  
\end{equation*}
\end{proof}
Lemmas \ref{converse-part} and \ref{direct-part} give the following:
\begin{corollary}\label{limit}
For $a<\derleft{\psi}(1)$ and $S_{n,a}:=\{e^{-na}\D{\rho}_n-\D{\sigma}_n>0\}$ we have
\begin{align*}
\lim_{n\rightarrow\infty}\oneover{n}\log\beta_n(S_{n,a})= -\phi(a).
\end{align*}
\end{corollary}
\begin{remark}\label{remark:converse-part}
Recall that $\vfi$ is strictly increasing on the interval
$\left[\derright{\psi}(0),+\infty\right)$.
An obvious modification of the above proof then yields that for any
$a>\derright{\psi}(0)$ and any sequence of tests $\{T_n\}$ satisfying
\begin{align*}
\limsup_{n\rightarrow\infty}\oneover{n}\log\beta_n(T_n)\le -\phi(a)
\end{align*}
we have
\begin{align*}
\liminf_{n\rightarrow\infty}\oneover{n}\log\alpha_n(T_n)
\ge -\{\phi(a)-a\}.
\end{align*}
In particular,
\begin{equation*}
\lim_{n\rightarrow\infty}\oneover{n}\log\alpha_n(S_{n,a})= -\{\phi(a)-a\}
\end{equation*}
for any $a>\derright{\psi}(0)$.
\end{remark}

\bigskip
Note that $\derright{\psi}(0)\le\derleft{\psi}(1)$ due to the convexity of $\psi$, and
the interval
\begin{equation*}
\I{\psi}:=\{a\in\R\,:\,\derright{\psi}(0)<a<\derleft{\psi}(1)\}
\end{equation*}
is nonempty if and only if the graph of $\psi$ is not a straight line segment. Corollary \ref{limit} and Remark
\ref{remark:converse-part} give the following:

\begin{theorem}\label{cor:convergence}
For $a\in\I{\psi}$ we have
\begin{align*}
\lim_{n\rightarrow\infty}
\oneover{n}\log\alpha_n(S_{n,a})
&=-\{\phi(a)-a\},
\\
\lim_{n\rightarrow\infty}
\oneover{n}\log\beta_n(S_{n,a})
&=-\phi(a).
\end{align*}
\end{theorem}
 \bigskip
 
\begin{figure}[H]
\begin{center}
\includegraphics*[width=0.9\textwidth]{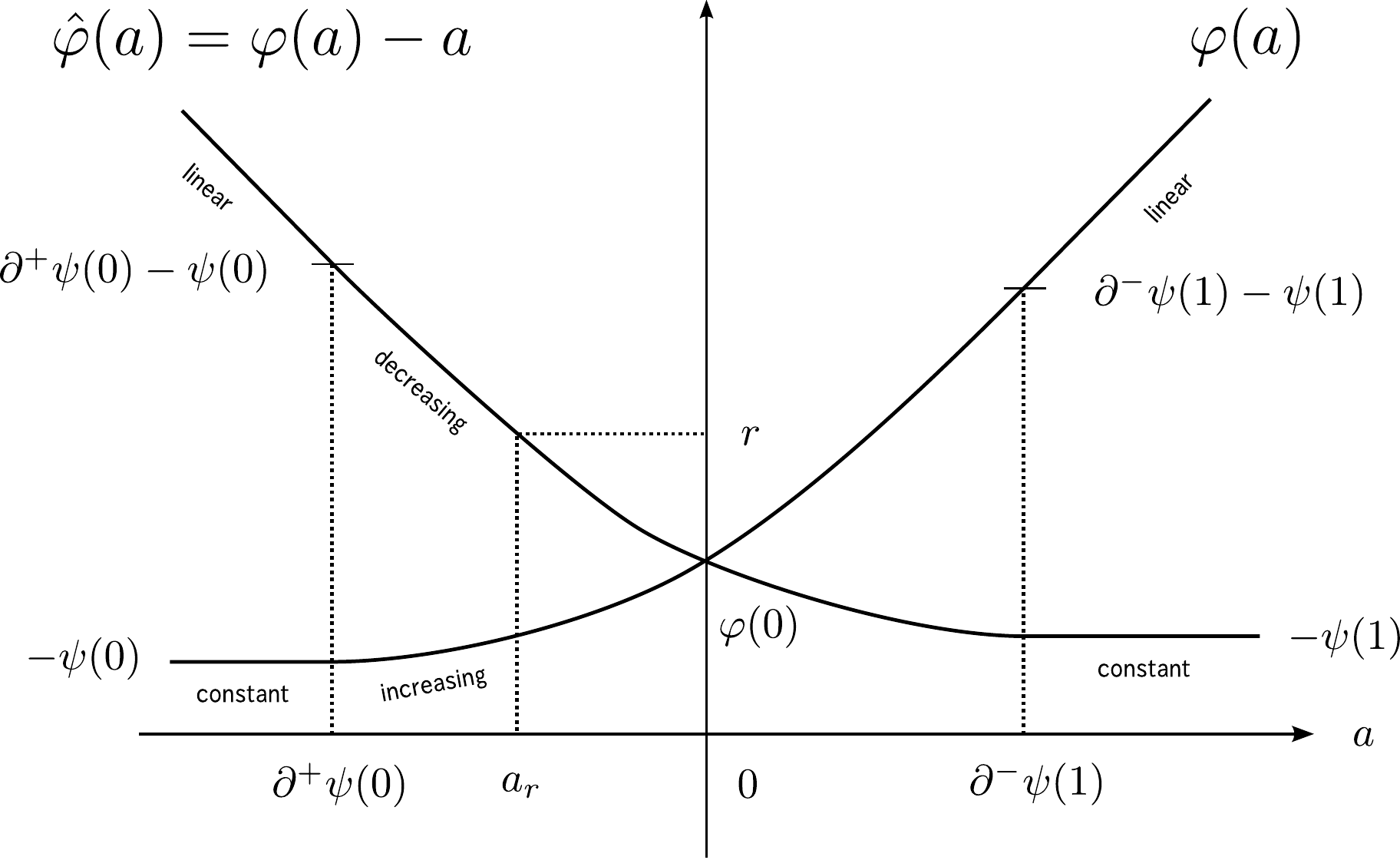}
\end{center}
\caption{the graphs of $\varphi$ and $\hat\varphi$ in a typical case}
\label{fig:phi}
\end{figure}

Now we are in a position to prove our main result.  
\begin{theorem}\label{thm:hoeffding bound}
For any $r\in\R$ we have
\begin{equation*}
\hli{r}=\hls{r}=\hlim{r}=-\sup_{0\le s<1}\frac{-sr-\psi(s)}{1-s}\,.
\end{equation*}
For $r<-\psi(1)$ all the above quantities are equal to $-\infty$; for $r\ge -\psi(1)$
we have
\begin{equation*}
\hlim{r}=-\max_{0\le s<1}\frac{-sr-\psi(s)}{1-s}=-\vfi(a_r)\,,
\end{equation*}
where $a_r$ is a unique number in $(-\infty,\derleft{\psi}(1)]$ such that
$\hat\vfi(a_r)=r$.
\end{theorem}
\begin{proof}
First assume that $r<-\psi(1)$.  
A straightforward computation shows that $\frac{-sr-\psi(s)}{1-s}$ tends to $+\infty$
as $s\nearrow 1$, hence $\sup_{0\le s<1}\frac{-sr-\psi(s)}{1-s}=+\infty$. Let
$T_n:=I-\supp\D{\sigma}_n$. Then
$\frac{1}{n}\log\alpha_n(T_n)=\psi_n(1)$ by definition, hence
$\limsup_n \frac{1}{n}\log\alpha_n(T_n)=\psi(1)<-r$ and obviously $\beta_n(T_n)=0$ for
all $n$. Hence (by using the convention $\log 0:=-\infty$) we get 
\begin{equation*}
\hlim{r} \le -\infty 
=-\sup_{0\le s<1}\frac{-sr-\psi(s)}{1-s}\,,
\end{equation*}
and the inequalities in \eqref{inequality:exponents} give the desired statement.
 
Now if $r\ge-\psi(1)$ then the 
properties of $\hat\vfi$ listed in (ii) of Lemma \ref{lem:phi} guarantee the existence
of a unique $a_r\le\derleft{\psi}(1)$ such that $\hat\vfi(a_r)=r$ (see Figure \ref{fig:phi} in \ref{graphs} for an illustration). Note that if $\derright{\psi}(1)=+\infty$ then $a_r=+\infty$ and we use the conventions of Remark \ref{infinity}. If
$\limsup_n\frac{1}{n}\log\alpha_n(T_n)<-r$ then there exists a $b<a_r$ such that for
all $b<a<a_r$ we have
\begin{equation*}
\limsup_n\frac{1}{n}\log\alpha_n(T_n)\le -\{\vfi(a)-a\}<-\{\vfi(a_r)-a_r\}=-r\,.
\end{equation*}   
By Lemma \ref{converse-part} 
\begin{equation*}
\liminf_n\oneover{n}\log\beta_n(T_n)\ge -\phi(a)\,,
\end{equation*}
and by taking the limit $a\nearrow a_r$ we obtain
\begin{equation*}
\liminf_n\oneover{n}\log\beta_n(T_n)\ge -\phi(a_r)\,.
\end{equation*}
Hence
\begin{equation*}
\hli{r}\ge -\phi(a_r)\,.
\end{equation*}
On the other hand, by Lemma \ref{direct-part} and Corollary \ref{limit} 
we have for any $a< a_r$
\begin{eqnarray*}
\limsup_n\frac{1}{n}\log\alpha_n(S_{n,a})&\le&-\{\vfi(a)-a\}<-\{\vfi(a_r)-a_r\}=-r\,,\\
\lim\frac{1}{n}\log\beta_n(S_{n,a})&=&-\vfi(a)\,,
\end{eqnarray*}
hence $\hlim{r}\le -\vfi(a)$. Now taking $a\nearrow a_r$ we get
\begin{equation*}
\hlim{r}\le -\phi(a_r)\,.
\end{equation*}
Taking the inequalities in \eqref{inequality:exponents} into account, we have 
\begin{equation*}
-\vfi(a_r)\le\hli{r}\le\hls{r}\le\hlim{r}\le-\vfi(a_r)\,.
\end{equation*}

To prove the last identity assume first that $r>-\psi(1)$. Then $a_r<\derleft{\psi}(1)$
and $s_r<1$, where $s_r:=\argmax_{0\le s\le 1}\{a_rs-\psi(s)\}$. Thus we have
\begin{align*}
r =\phi(a_r)-a_r
=a_rs_r-\psi(s_r)-a_r
\ge a_rs-\psi(s)-a_r\,,\ds\ds\ds 0\le s\le 1\,,
\end{align*}
hence
$a_r \ge \frac{-r-\psi(s)}{1-s}$ for any $0\le s<1$ with equality for $s=s_r$. Then
\begin{align}
\phi(a_r)
=a_rs_r-\psi(s_r)
\ge a_rs-\psi(s)\ge\frac{-sr-\psi(s)}{1-s}\,,\ds\ds\ds 0\le s<1\,,
\label{another2}
\end{align}
and equality holds for $s=s_r$.

Now if $r=-\psi(1)$ then $a_r=\derleft{\psi}(1)$ and
$\vfi(a_r)=\derleft{\psi}(1)-\psi(1)$, and one can easily see that
\begin{equation*}
\lim_{s\nearrow 1}\frac{-s(-\psi(1))-\psi(s)}{1-s}=\derleft{\psi}(1)-\psi(1)\,.
\end{equation*}
If $\derleft{\psi}(1)=+\infty$ then this gives the desired identity immediately. If $\derleft{\psi}(1)<+\infty$ then
the statement follows from the fact that inequalities in \eqref{another2} are still
valid (with $s_r=1$). 
\end{proof}
\smallskip

One can get a certain interpolation between the setting of Stein's lemma and the theorem
for the Hoeffding bound by requiring that the $\alpha_n$'s converge to zero
exponentially, but without constraint on the value of the exponent. The corresponding
exponents for the $\beta_n$'s are $\hli{0},\s\hls{0}$ and $\hlim{0}$. We have the
following:
\begin{proposition}\label{prop:Stein exp}
If $\psi(1)=0$ then
\begin{equation*}
\hli{0}=\hls{0}=\hlim{0}=-\derleft{\psi}(1)\,.
\end{equation*}
Moreover, if $\rho$ and $\sigma$ satisfy the upper factorization property and
$\supp\D{\rho}_n\le\supp\D{\sigma}_n$ for all $n\in\N$ then
\begin{equation*}
\hlim{0}\ge -\srm{\rho}{\sigma}\,.
\end{equation*}
\end{proposition}
\begin{proof}
The first statement is a special case of Theorem \ref{thm:hoeffding bound}. To see this,
take $r=-\psi(1)=0$; then 
$a_r=\derleft{\psi}(1)$ and $\vfi(a_r)=\derleft{\psi}(1)$. The second statement follows
from \eqref{derleft-relentr}.
\end{proof}
\smallskip

In studying Stein's lemma, one is interested in the exponents
$B(\vec\rho\,||\,\vec\sigma),\s \underline{B}(\vec\rho\,||\,\vec\sigma)$ and
$\overline{B}(\vec\rho\,||\,\vec\sigma)$, where
\begin{equation*}
B(\vec\rho\,||\,\vec\sigma):=\inf_{\{T_n\}}\biggl\{
\lim_{n\rightarrow\infty} \oneover{n}\log\beta_n(T_n)
\biggm|
\lim_{n\rightarrow\infty} \alpha_n(T_n)=0
\biggr\} \,,
\end{equation*}
and $\underline{B}(\vec\rho\,||\,\vec\sigma)$ and
$\overline{B}(\vec\rho\,||\,\vec\sigma)$ are defined similarly, by taking
$\liminf_{n\rightarrow\infty} \oneover{n}\log\beta_n(T_n)$ and
$\limsup_{n\rightarrow\infty} \oneover{n}\log\beta_n(T_n)$. Obviously,
$\underline{B}(\vec\rho\,||\,\vec\sigma)\le \overline{B}(\vec\rho\,||\,\vec\sigma)
\le B(\vec\rho\,||\,\vec\sigma)$, and Theorem \ref{thm:hoeffding bound} has the
following consequence:
\begin{proposition}\label{Stein}
If $\psi(1)<0$ then
\begin{equation*}
\underline{B}(\vec\rho\,||\,\vec\sigma)=
 \overline{B}(\vec\rho\,||\,\vec\sigma)=
 B(\vec\rho\,||\,\vec\sigma)=-\infty =-\srm{\rho}{\sigma}\,.
\end{equation*} 
If $\psi(1)=0$ then 
\begin{equation*}
B(\vec\rho\,||\,\vec\sigma) \le -\derleft{\psi}(1)\,.
\end{equation*}
\end{proposition}
\begin{proof}
If $\psi(1)<0$ then there exists an $N\in\N$ such that
$\Tr \D{\rho}_n [\supp \D{\sigma}_n]<1$ for all $n\ge N$,
i.e.\ $\supp\D{\rho}_n\nleq \supp \D{\sigma}_n$ and hence
$\sr{\rho_n}{\sigma_n}=+\infty$ for all $n\ge N$, implying
$\srm{\rho}{\sigma}=+\infty$. The rest of the statements follow immediately from 
the fact that $B(\vec\rho\,||\,\vec\sigma)\le\hlim{0}$, and that $\hlim{0}=-\infty$
when $\psi(1)<0$ by Theorem \ref{thm:hoeffding bound} and $\hlim{0}=-\derleft{\psi}(1)$
when $\psi(1)=0$ by Proposition \ref{prop:Stein exp}.
\end{proof}
Note that when $\rho=\rho_1^{\otimes\infty}$ and $\sigma=\sigma_1^{\otimes\infty}$ are
product states with $\supp\D{\rho}_1\le\supp\D{\sigma}_1$ then $\psi(1)=0$
and $\derleft{\psi}(1)=\sr{\rho_1}{\sigma_1}$, and we get back the well-known formula
for the direct part of Stein's lemma.

\section{Concluding remarks}

We have studied various error exponents, including the Chernoff and Hoeffding bounds in a binary (asymptotic) hypothesis testing problem. 
While following a rather general formulation, the main applicability of our results is the hypothesis testing problem for two states on a spin chain, both satisfying the factorization properties given in Definition \ref{def:factorization}. That the study of such states is sufficiently well motivated was established in \cite{HMO}, where we have shown that the factorization properties are satisfied by the global Gibbs states of translation-invariant finite-range interactions. Other important examples for states 
to which our results may be applicable are provided by the Markov-type class of finitely correlated states \cite{FNW} (see e.g.~Example \ref{example 1} and \cite{HMO}). 
While finitely correlated states always satisfy the upper factorization property \cite{HMO}, it is an open question at the moment to find necessary and sufficient conditions for the lower factorization property to hold. As Remark \ref{separation} suggests, it may be possible to prove the validity of assumptions \ref{existence} and \ref{diff2} for finitely correlated states even if the lower factorization property fails to hold. Similar conditions to our factorization properties were used at various places in the literature. Probably the closest to our factorization properties is the $^*$-mixing condition (see e.g.~\cite{Bjelakovic} and references therein). However, the relation among these conditions and the factorization properties in the quantum setting is an open question at the moment.

Our main tool in deriving the upper bounds in Lemma \ref{direct-part} was the powerful trace inequality \eqref{trace inequality} discovered in \cite{Aud},
that was successfully applied to give the Chernoff \cite{Aud} and Hoeffding
\cite{Hayashi} upper bounds in the i.i.d.~case. Our Corollary \ref{cor direct-part} is
an extension of \cite[Theorem 1]{Hayashi} to the very general setting of Section
\ref{error exponents}, and a slight simplification as well, as the tests only depend on
the parameter $a$, and not on $s$ as in \cite{Hayashi}.

The main idea in deriving the lower bound \eqref{liminf-ineq} in Section \ref{converse-part 1} is to relate the quantum problem to a
classical hypothesis testing problem by the method of \cite{NSz} and then use large
deviation techniques to treat the classical problem. This approach was used in the
i.i.d.\ case to derive the quantum Chernoff \cite{NSz} and Hoeffding \cite{Nagaoka}
lower bounds. In \cite{Nagaoka} the two states were implicitly assumed to have the same
support, which assumption 
can easily be removed by restricting the classical probability distributions onto the
intersection of their supports; this approach was followed in \cite{ANSzV}. In Section
\ref{converse-part 1} we have followed a different way to circumvent the restriction of
equivalent supports, by slightly modifying the way to assign classical measures to the
original states. 
In the non-i.i.d.~case it is a natural choice to use the G\"artner-Ellis theorem to establish the lower bound in the classical hypothesis testing problem, and 
the differentiability condition in assumption \ref{diff2} is essentially the requirement of the differentiability of the logarithmic moment generating function in the G\"artner-Ellis theorem. As we argue in Section \ref{some remarks}, the main results of Section \ref{lower bounds}, Theorems \ref{lower bound app} and \ref{Nagaoka-bound} are essentially giving the theorem for the Chernoff bound in an appropriate setting. The fact that Corollary \ref{iid lower bound} is true for all real numbers $a$ seems to be new even in the i.i.d.~setting. The exclusion of the cases $a=\derright{\psi}(0)$ and $a=\derleft{\psi}(1)$ in Theorem \ref{lower bound app} is strongly related to the possibility of a pathological situation when the graph of $\psi$ becomes a straight line, and could possibly be removed if a similar characterization to that in Lemma \ref{affine structure} was available also in the non-i.i.d.~case.

It is well-known in the information spectrum approach that the limits of the quantities $\frac{1}{n}\log \alpha_n(S_{n,a})$ and $\frac{1}{n}\log \beta_n(S_{n,a})$ are strongly related to the theorem for the Hoeffding bound, as was emphasized e.g.~in \cite{Han} and \cite{NH}. In Lemma \ref{converse-part} we follow the way of \cite{Nagaoka} to derive these limits from the Chernoff-type theorems of Section \ref{lower bounds}. Theorem \ref{cor:convergence} was stated as a conjecture in \cite{Nagaoka} and it was proven shortly after in \cite{Nagaoka-2} in the i.i.d.~setting for the values of $a$ between $-\srm{\sigma_1}{\rho_1}$ and $\srm{\rho_1}{\sigma_1}$; this coincides with our $\I{\psi}$ when $\supp\D{\rho}_1=\supp\D{\sigma}_1$ is assumed in the i.i.d.~setting. The importance of the above limits are clearly shown by the fact that the results of Theorem \ref{thm:hoeffding bound} and Propositions \ref{prop:Stein exp} and \ref{Stein} hold true whenever Corollary \ref{limit} is true (here we benefit from the fact that Lemma \ref{direct-part} is unconditionally true in the most general setting, showing again the power of inequality \eqref{trace inequality}).

The interpretation of $Q(\rho,\,\sigma):=\min_{0\le s\le 1}\Tr\D{\rho}^s\D{\sigma}^{1-s}$ as a distinguishability measure on the state space of a finite dimensional quantum system was investigated in \cite{Aud}, where a detailed analysis of its properties and its relation to other measures (like fidelity, trace distance and relative entropy) was given. Here we would like to stress the importance of its negative logarithmic version
 \begin{equation*}
 \cb{\rho}{\sigma}:=-\min_{0\le s\le 1}\log\Tr\D{\rho}^s\D{\sigma}^{1-s}=-\log Q(\rho,\,\sigma)\,.
 \end{equation*}
 It is jointly convex in its variables (due to Lieb's concavity theorem),
 monotonic decreasing under $2$-positive trace-preserving maps \cite{Petz,Uhlmann}, and easily seen to be strictly positive, thus 
 sharing some of the most important properties of relative entropy. Moreover, if $\vec{\rho},\vec{\sigma}\in\factstates{\C}$ then the uniform convergence established in Lemma \ref{factorization limit} shows that the limit
 \begin{equation}\label{mean chernoff}
 \cbm{\rho}{\sigma}:=\lim_{n\to\infty}\frac{1}{n} \cb{\rho_n}{\sigma_n}
 \end{equation}
 exists (and coincides with $\vfi(0)$), further extending the analogy with the relative entropy. Theorem \ref{Nagaoka-bound} for the Chernoff bound gives that 
 \begin{equation*}
 \lim_{n\to\infty}\frac{1}{n}\log\min_{0\le T_n\le I}\{\alpha_n(T_n)+\beta_n(T_n)\}=-\cbm{\rho}{\sigma},
 \end{equation*}
 thus giving an operational interpretation to the mean Chernoff distance, and showing that it plays exactly the same role in the symmetric setting of the theorem for the Chernoff bound as the mean relative entropy plays in the asymmetric setting of Stein's lemma. Obviously, the asymptotic quantity \eqref{mean chernoff} is still jointly concave and monotonic decreasing under $2$-positive trace-preserving maps; it is not clear, however, whether the strict positivity property is preserved under taking the limit.


\section*{Acknowledgments}

Partial funding by PRESTO "Quanta and Information" in JST (T.O.),
Grant-in-Aid for Scientific Research (B)17340043 (F.H.) and Grant-in-Aid for JSPS
Fellows 18\,$\cdot$\,06916, as well as the JSPS fellowship P06916 (M.M.) are gratefully
acknowledged.

\appendix
\swapnumbers

\def\thesection{Appendix \Alph{section}} 
\section{}\label{min}

\def\thesection{\Alph{section}} 

Note that Lemma \ref{Audenaert} becomes trivial when $\B(\iH)$ is replaced with a
commutative $C^*$-algebra. Indeed, in this case elements of the algebra are functions
on some compact space $X$, and if $f,g$ are non-negative functions then 
\begin{equation*}
\frac{1}{2}\bz f(x)+g(x)-|f(x)-g(x)|\jz=\min\{f(x),g(x)\}
\le f(x)^{s}g(x)^{1-s}\,,\ds\ds\ds s\in[0,1]\,,
\end{equation*}
and the trace can be replaced with integration with respect to an arbitrary positive
measure. The minimum of two nonnegative operators $A,B\in\B(\iH)$ might
be defined as the unique self-adjoint $X\in\B(\iH)$ for which   
\begin{equation*}
X\le A,\ds X\le B\ds\ds\text{and}\ds\ds Y\le A,\ds Y\le B\THEN Y\le X
\end{equation*}
holds. Note, however, that such an operator does not exist in general, even when $A$ and $B$ commute with each other.
On the other hand, the following is true:
\begin{proposition}
For any nonnegative operators $A,B\in\B(\iH)$, we have
\begin{align*}
\max\{\Tr X\,:\, X\le A,\s X\le B\} \le \Tr A^{s}B^{1-s}\,,\ds\ds\ds
s\in [0,1]\,.
\end{align*}
\end{proposition}
The proof follows immediately from Lemma \ref{Audenaert} and the following lemma,
which is a special case of
the duality theorem in multiple hypothesis testing \cite{Yuen,Holevo}.
Since the proof of the binary case is immediate, we include it
for readers' convenience.
\begin{lemma}
For any nonnegative operators $A,B\in\B(\iH)$, we have
\begin{eqnarray*}
\max\{\Tr X\,:\, X\le A,\s X\le B\}&=&\min_{0\le T\le I}\{\Tr A(I-T)+\Tr BT \}\\
&=&\Tr \half\bz A+B-|A-B|\jz\,.
 \end{eqnarray*}
\end{lemma}
\begin{proof}
Suppose that $X\le A$, $X\le B$.
Then $X=X^*$
and for any operator $T$ satisfying $0\le T\le I$ we have
\begin{align*}
\Tr X = \Tr X(I-T)+\Tr XT
\le \Tr A(I-T)+\Tr BT.
\end{align*}
Conversely, let $S:=\{A-B>0\}$ and  $X:=A(I-S)+BS=\half\bz A+B-|A-B|\jz$. Then
\begin{equation*}
X=A-(A-B)_+\le A,
\ds\ds\ds
X=B-(A-B)_-\le B,
\end{equation*}
where $(A-B)_{+}$ and $(A-B)_-$ denote the positive and the negative parts of $A-B$,
respectively.
\end{proof}
Loosely speaking, the above shows that as long as $\Tr$ is taken,
$\half\bz A+B-|A-B|\jz$ can be considered as the minimum of $A$ and $B$. Note, however,
that $\half\bz A+B-|A-B|\jz$ need not even be positive semidefinite when both $A$ and
$B$ are positive semidefinite. A simple counterexample is given by $A:=\proj{a}$ and
$B:=\proj{b}$ with $a=(1,i)$ and $b=(1,1)$. 
  
\def\thesection{Appendix \Alph{section}} 
\section{}\label{differentiability}

\def\thesection{\Alph{section}} 

Unfortunately, assumptions \ref{existence} and \ref{diff2} in Section \ref{converse-part 1}
seem to be difficult to verify in general correlated cases. Below we show a specific
example for a pair of correlated states on a spin chain for which assumptions \ref{existence}
and \ref{diff2} can directly be verified. Note that the example is non-classical in the
sense that the local densities $\D{\rho}_n$ and $\D{\sigma}_n$ need not commute with
each other. However, both $\rho$ and $\sigma$ exhibit only classical correlations among
the sites of the chain, i.e.~all local densities $\D{\rho}_n,\s\D{\sigma}_n,\s n\in\N$,
are separable.
 
\begin{example}\label{example 1}
Let $T=\{T_{xy}\}_{x,y\in\X}$ and $S=\{S_{xy}\}_{x,y\in\X}$ be irreducible stochastic
matrices with corresponding faithful stationary distributions $r=\{r_x\}_{x\in\X}$ and
$p=\{p_x\}_{x\in\X}$ on some finite set $\X$. Moreover, let $\{\state_{xy}\}_{x,y\in\X}$
and $\{\vfi_{xy}\}_{x,y\in\X}$ be sets of states on a finite-dimensional $C^*$-algebra
$\A$ and $\Theta_x:=\sum_y T_{xy}\state_{xy},\s \Phi_x:=\sum_y S_{xy}\vfi_{xy}$. The
local states
\begin{eqnarray*}
  \rho_n&:=&\sum_{x_1,\ldots, x_{n}\in\X}\,r_{x_1} \bz T_{x_1x_2}\state_{x_1\,x_2}\jz \otimes\ldots\otimes \bz T_{x_{n-1}x_n}\state_{x_{n-1}\,x_{n}}\jz\otimes\Theta_{x_n}\,,\\
  \sigma_n&:=&\sum_{x_1,\ldots, x_{n}\in\X}\,p_{x_1} \bz S_{x_1x_2}\vfi_{x_1\,x_2}\jz \otimes\ldots\otimes \bz S_{x_{n-1}x_n}\vfi_{x_{n-1}\,x_{n}}\jz\otimes\Phi_{x_n}\,,
\end{eqnarray*}
are easily seen to extend to translation-invariant states $\rho$ and $\sigma$ on the
spin chain $\C=\otimes_{k\in\Z}\A$. (Actually, $\rho$ and $\sigma$ are ergodic finitely
correlated states with a commutative auxiliary algebra in their generating triples; see
\cite{FNW} and also \cite{HMO} for details.)

Let us assume that there exists a set of non-zero projections $\{P_x\}_{\in\X}$ in $\A$
with orthogonal ranges such that
\begin{equation}\label{condition1}
\supp\D{\state}_{xy}\vee\supp\D{\vfi}_{xy}\le P_x\ds\ds \text{and also}\ds\ds
\supp\D{\state}_{xy}\wedge\supp\D{\vfi}_{xy}\ne 0\,,\ds x,y\in\X.
\end{equation}
Then
\begin{eqnarray*}
\Tr\D{\rho}_n^{s}\,\D{\sigma}_n^{1-s}&=&\sum_{x_1,\ldots,x_{n}}
r_{x_1}^{s}p_{x_1}^{1-s}\,\bz\Tr\D{\Theta}_{x_n}^{s}\D{\Phi}_{x_n}^{1-s}
\jz\prod_{k=1}^{n-1}\,\bz T_{x_k x_{k+1}}^{s}S_{x_k x_{k+1}}^{1-s}
\Tr\D{\state}_{x_k x_{k+1}}^{s}\D{\vfi}_{x_k x_{k+1}}^{1-s} \jz\\
&=&
\inner{a(s)}{Q(s)^{n-1}b(s)}
\end{eqnarray*}{\color{black}}
for every $s\in\R$ with 
\begin{equation*} 
a(s)_{x}:=r_{x}^{s}p_{x}^{1-s},\ds\ds b(s)_{x}:=\Tr\D{\Theta}_{x}^{s}\D{\Phi}_{x}^{1-s}\ds\ds\text{and}\ds\ds 
Q(s)_{x,y}:=T_{x y}^{s}S_{xy}^{1-s}\Tr\D{\state}_{x y}^{s}\D{\vfi}_{xy}^{1-s}\,.
\end{equation*}
Now if $Q(s)$ is irreducible for some (and hence for all) $s\in\R$ then by the
Perron-Frobenius theorem we have
\begin{equation*}
\psi(s)=\lim_{n\to\infty}\frac{1}{n}\log \inner{a(s)}{Q(s)^{n-1}b(s)} =\log r(s)\,,\ds\ds\ds s\in\R\,,
\end{equation*}
where $r(s)$ is the spectral radius of $Q(s)$ (see e.g.\ \cite[Theorem 3.1.1]{DZ}).
Being a simple eigenvalue, the function $s\mapsto r(s)$ is smooth (cf.~\cite{Kato}),
and so is $\psi$, hence assumptions \ref{existence} and \ref{diff2}  are
satisfied in this case.
 
Assume now that $\supp\D{\rho}_n\le\supp\D{\sigma}_n$ for some $n\ge 2$. This is easily
seen to be equivalent to the conditions
\begin{eqnarray}
T_{xy}=0 \ds&\text{if}&\ds S_{xy}=0\,, \label{condition2}\\
\supp\D{\state}_{xy}\le\supp\D{\vfi}_{xy}\ds&\text{if}& \ds T_{xy}>0\,,\label{condition3}
\end{eqnarray}
and hence is independent of the value of $n$. (Note that the first condition states
that the classical Markov chain generated by $T$ and $r$ is absolutely continuous with
respect to that generated by $S$ and $p$.) It is easily seen that in this case $Q(s)$
is irreducible for every $s\in\R$, hence we can apply the above argument and obtain
$\psi(s)=\log r(s),\s s\in\R$. Simplicity of $r(s)$ as an eigenvalue of $Q(s)$ yields
that one can choose the corresponding Perron-Frobenius
(left) eigenvectors $\xi(s)=\{\xi_x(s)\}_{x\in\X}$ to form a strictly positive
probability distribution for all $s\in\R$ such that the function $s\mapsto\xi(s)$ is
smooth (cf.~\cite{Kato}).
 Let $e:=(1,\dots,1)$ be the identity
vector. Using the facts that 
$r(s)=\inner{\xi(s)Q(s)}{e}$, that $\inner{\xi'(s)}{e}=0$ due to the fact that
$\inner{\xi(s)}{e}=1$ for all $s$, and that $Q(1)=T,\s \xi(1)=r,\s r(T)=1$, we obtain
\begin{align*}
\psi'(1)&=\frac{r'(1)}{r(1)}
=\inner{\xi'(1)}{Q(1)e}+\inner{\xi(1)}{Q'(1)e} \\
&=\sum_x r_x\sr{T_{x.}}{S_{x.}}+\sum_{xy}r_xT_{xy}\sr{\state_{xy}}{\vfi_{xy}}\,.
\end{align*}
A straightforward computation then shows that the latter expression is exactly the
mean relative entropy $\srm{\rho}{\sigma}$. Similarly, if we impose the condition
$\supp\D{\rho}_n\ge\supp\D{\sigma}_n$ for some (and hence for all) $n\ge 2$ then we
obtain $\psi'(0)=-\srm{\sigma}{\rho}$.
\end{example}

\begin{remark}\label{separation}
In the above construction, let $\A$ be isomorphic to the function algebra on some
finite set $\X$, and let  $\D{\state}_{xy}:=\D{\vfi}_{xy}:=\egy_{\{x\}}$
(the indicator function of $\{x\}$) for all $x,y\in\X$. Then $\D{\rho}_n$ and
$\D{\sigma}_n$ are the densities of the $n$-site restrictions of the Markov measures $\mu$ and $\nu$
generated by  $(T,r)$ and $(S,p)$, respectively, and the conditions  
\eqref{condition1} and \eqref{condition3} are automatically
satisfied.

Now it is easy to see that $\mu$ and $\nu$ satisfy the lower factorization property if and only $T$ and $R$ are entrywise strictly positive matrices, which condition is sufficient but not necessary for \eqref{condition2} to hold.
Thus the above construction provides examples for situations when the lower factorization property is not satisfied while assumptions \ref{existence} and \ref{diff2} hold true.
\end{remark}

\begin{example}\label{example 2}
Let $\rho$ and $\sigma$ be the global Gibbs states of translation-invariant finite-range
interactions $\Phi$ and $\Psi$, respectively. The local Gibbs state $\rho_n^G$ for
$\Phi$ has the density $e^{-H_n(\Phi)}/\Tr e^{-H_n(\Phi)}$, where $H_n(\Phi)$ is the
local Hamiltonian of $\Phi$ inside $[1,n]$; the local Gibbs state $\sigma_n^G$ is
defined similarly for $\Psi$.
Since there is a constant $\lambda\ge 1$ (independent of $n$) such that
$\lambda^{-1}\rho_n\le\rho_n^G\le\lambda\rho_n$ and
$\lambda^{-1}\sigma_n\le\sigma_n^G\le\lambda\sigma_n$ (see \cite[Lemma 2.1]{HMOP}),
the function $\psi$ is written as
\begin{equation*}
\psi(s)=\lim_{n\to\infty}{1\over n}\log
\mathrm{Tr}\,e^{-sH_n(\Phi)}e^{-(1-s)H_n(\Psi)}
-sP(\Phi)-(1-s)P(\Psi),\qquad s\in\R,
\end{equation*}
where $P(\Phi):=\lim_{n\to\infty}{1\over n}\log\Tr e^{-H_n(\Phi)}$, the
pressure of $\Phi$. We have $\psi(0)=\psi(1)=0$, and the Golden-Thompson trace
inequality yields
\begin{equation*}
\psi(s)\ge\xi(s):=P(s\Phi+(1-s)\Psi)-sP(\Phi)-(1-s)P(\Psi),\qquad s\in\R.
\end{equation*}
Here equality cannot hold in general as is immediately seen in the case of product
states. Note that $\xi(s)$ corresponds to one of the candidates proposed in
\cite{OH} to obtain the Quantum Hoeffding bound.
By \cite[Theorem 2.4 and Lemma 2.3]{HMOP} we notice that $\xi$ is differentiable on
$\R$ and moreover
\begin{eqnarray*}
\xi'(0)&=&\partial P_\Psi(\Phi-\Psi)-P(\Phi)+P(\Psi)\\ 
&=&-\sigma(A_{\Phi-\Psi})-P(\Phi)+s(\sigma)-\sigma(A_\Psi) \\
&=&s(\sigma)-\sigma(A_\Phi)-P(\Phi) \\
&=&-S_M(\sigma||\rho),
\end{eqnarray*}
where $A_\Psi$ is the mean energy of $\Psi$ and $s(\sigma)$ is the mean entropy of
$\sigma$. Similarly $\xi'(1)=S_M(\rho||\sigma)$. Hence we have
$\psi'(0)=-S_M(\sigma||\rho)$ and $\psi'(1)=S_M(\rho||\sigma)$ as long as $\psi$ is
differentiable at $0,1$. In particular, when $H_n(\Phi)$ and $H_n(\Psi)$
commute for all $n$, it is obvious that $\psi=\xi$. But this situation is essentially
classical since for all $n$ the densities of $\rho_n$ and $\sigma_n$ commute, too.
\end{example}


\def\thesection{Appendix \Alph{section}} 
\section{}\label{graphs}

\def\thesection{\Alph{section}} 

\s\newline

\begin{figure}[H]
\begin{center}
\includegraphics*[width=0.7\textwidth]{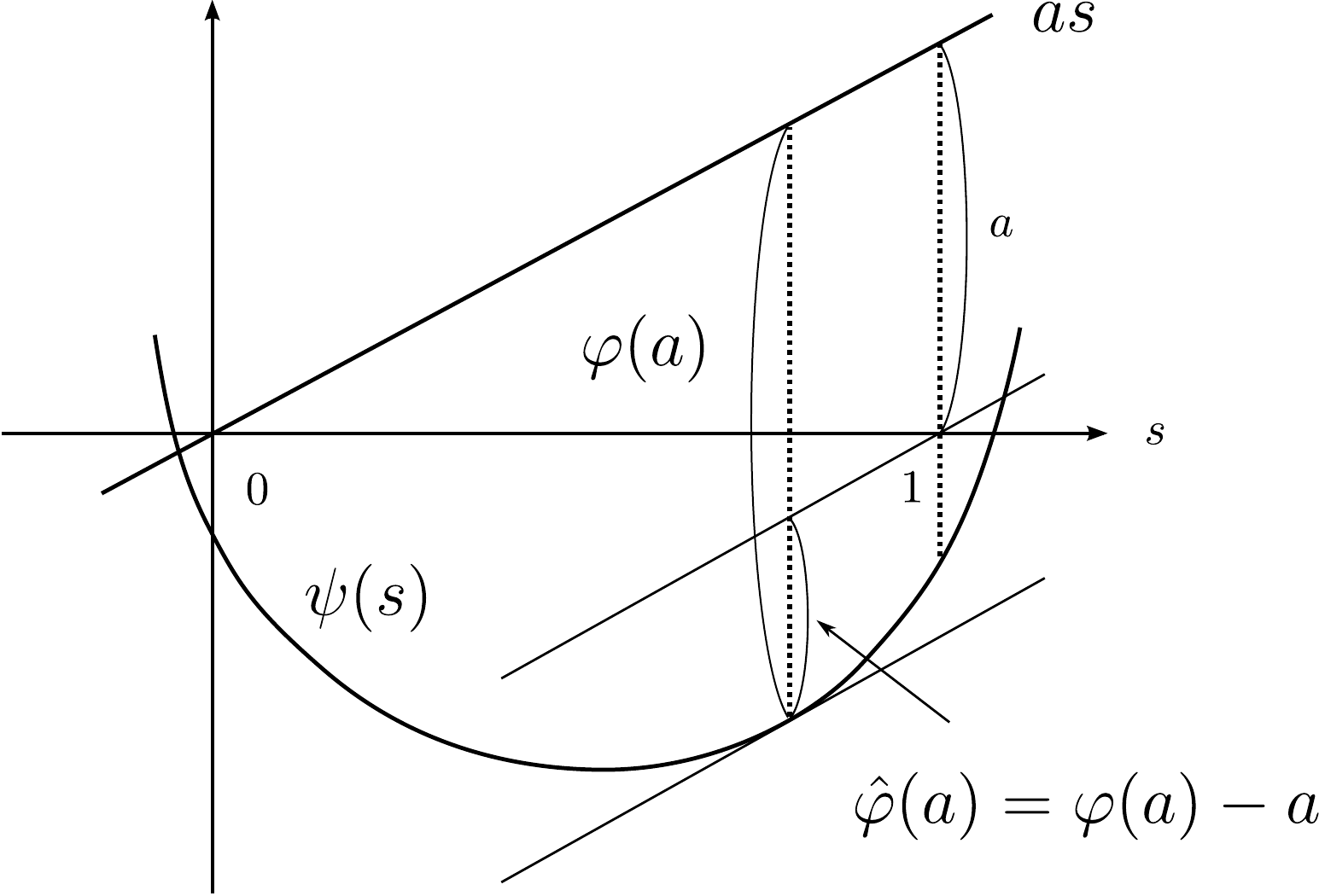}
\end{center}
\caption{the definitions of $\vfi$ and $\hat\vfi$ with a typical $\psi$}
\label{fig:psi}
\end{figure}

\vspace{1.5cm}

\vspace{1cm}
 \s

\end{document}